\documentclass[aps,pre]{revtex4}
\usepackage{epsfig}
\usepackage{amsmath,amssymb}
\begin{document}
\title{Supported bilayers: combined specular and diffuse x-ray scattering}
\author{L. Malaquin$^{1,2}$, T. Charitat$^1$ and J. Daillant$^2$}
\affiliation{$^1$Institut Charles Sadron, Universit\'e de Strasbourg, CNRS UPR 22\\ 
	23 Rue du Loess, BP 84047\\ 
	67034 Strasbourg cedex 2, France\\
$^2$CEA,IRAMIS,LIONS, CEA-Saclay, b\^at. 125, F-91191 Gif-sur-Yvette Cedex, France.}

\begin{abstract} 
A new method is proposed for the analysis of specular and off-specular reflectivity from supported lipid bilayers.  
Both thermal fluctuations and the ``static" roughness induced by the substrate are carefully taken into account.
Examples from supported bilayers and more complex systems comprising a bilayer adsorbed or grafted 
on the substrate and another ``floating" bilayer are given.
The combined analysis of specular and off-specular reflectivity allows the precise determination of the structure
of adsorbed and floating bilayers, their tension, bending rigidity and interaction potentials.
We show that this new method gives a unique opportunity to investigate phenomena like protusion modes of adsorbed bilayers
and opens the way to the investigation of more complex systems including different kinds of lipids, cholesterol or peptides.
\end{abstract}
\maketitle

\section{Introduction}
There is a great interest in determining the elastic properties and interactions between membranes 
at the microscopic scale. One attractive way to do it is to investigate their thermal fluctuations. 
This can be done for example by recording the shape of vesicles (LUVs, GUVs) as a function of time 
and careful image analysis \cite{pecreaux2004}. 
The wavelength of fluctuations which can be investigated by this method is however limited to $\approx 1 \mu m$. 
Shorter wavelengths can be investigated by using x-rays.
X-ray scattering has indeed been used to study unilamellar vesicles, and can be as precise as to demonstrate 
an asymmetry between the inner and outer leaflets \cite{Brzustowicz:ko5011}. 
Membrane fluctuations have however been mainly investigated in multilamellar vesicles 
\cite{petrache(pre1998),petrache(biophysj2004), NagleBBA2000}. The so-called Caill\'e theory \cite{caille72,Nall89} 
for x-ray diffraction from lyotropic liquid crystals enables in particular the determination of
 $\sqrt{\kappa B}$ where $\kappa$ is the bending rigidity of the membrane and B the stack compression modulus. 
Another advantage of the method \cite{rand(bba1989),NagleBBA2000} is that the sample can be osmotically compressed 
in a systematic way. 
The structure of the sample can then be determined as a function of osmotic pressure $\Pi$, 
resulting in particular in $\Pi(x)$ curves which can be fitted to interaction potential models, 
usually taking into account hydration forces, van der Waals forces and the so-called Helfrich entropic 
interaction whose characteristic parameters can be evaluated.\\
A significant progress was the use of well orientated multilamellar stacks, 
either spin-coated onto substrates \cite{Salditt2000} 
or freely suspended \cite{Jeu96} where simultaneous fitting of specular and off-specular scattering 
allows an independent determination of $\kappa$ and $B$. 
However, all these methods intrinsically suffer from the large number of defects in the sample, 
leading for example to inconsistencies between the temperature dependence of the elastic parameters and the observation 
of an unbinding transition \cite{Salditt2000}.\\
This issue can in principle be solved by looking at supported bilayers. 
When obtained by vesicle fusion \cite{brian(pnas1984)}, it is likely that some defects will remain, 
but almost defect-free samples can in principle be obtained by using the Langmuir-Blodgett (LB) technique 
\cite{charitat1999}. 
Their structure has been investigated in particular using atomic force microscopy \cite{Giocondi2004861} and
x-ray scattering. X-ray reflectivity has for example been used to determine the structure of model 
lipid rafts in microfluidics cells \cite{nickel08}, and x-ray diffraction on lipid bilayers adsorbed on a substrate in 
water has recently been demonstrated \cite{miller(PRL2008)}.\\
One disadvantage of adsorbed bilayers is that strongly adsorbed samples would be very far from natural conditions. 
For this reason, polymeric cushions have been used as spacers \cite{Wagner2000,Sinner2001}. 
Another possibility is to use a first bilayer as spacer, 
the second being less adsorbed on the substrate \cite{charitat1999}. 
A promising development is finally the investigation of bilayer membranes spanning microfabricated 
holes \cite{Beerlink2008} opening the possibility of investigating membranes in an asymmetric environment.\\
In this paper, we consider adsorbed bilayers where the spacer is either a first lipid bilayer 
adsorbed on the substrate (double bilayers) \cite{charitat1999} or a mixed 
octadecytrichlorosilane (OTS) layer - lipid monolayer (OTS bilayers), where the OTS layer is chemically grafted 
on the substrate and the lipid monolayer is deposited using the LB technique \cite{hughes2002}. 
A first study of such floating bilayers, mainly limited to the investigation of their structural and elastic 
properties has been recently published \cite{daillant2005} and further extended to the determination 
of the interaction potential. The aim of this paper is to present the underlying 
theory leading to the full determination of structural and elastic properties of bilayer membranes 
and their interaction potentials.\\
We extend to more complex samples, in particular double bilayers, a first analysis of the height-height 
correlation functions of correlated interfaces by Swain et al. \cite{andelman1999,andelman2001}. 
Fluctuation spectra of free and supported membrane pairs have also been calculated in \cite{merath07}.
We show in particular that only a joint analysis of the specular and off-specular reflectivities allows a 
full characterization of the system. 
Specular reflectivity is more sensitive to the adsorbed bilayer whereas off-specular scattering 
is more sensitive to the floating bilayer. 
It consists in the systematic convolution of the substrate height-height correlations 
with susceptibility functions in order to appropriately propagate the correlations, 
and also takes into account thermal fluctuations. 
We then perform an efficient numerical integration to obtain scattering cross-sections and intensities,
taking precisely into account resolution functions. 
This method is quite general and could be applied to similar problems like wetting films \cite{pershan(1991)}, 
polymer thin films or synthetic multilayers \cite{Daillant1999}.\\

\section{Supported bilayers}
\label{secgene}
\subsection{Free energy}
\label{subsectfree}
\begin{figure}[h]
\begin{center}
\includegraphics[width=10cm]{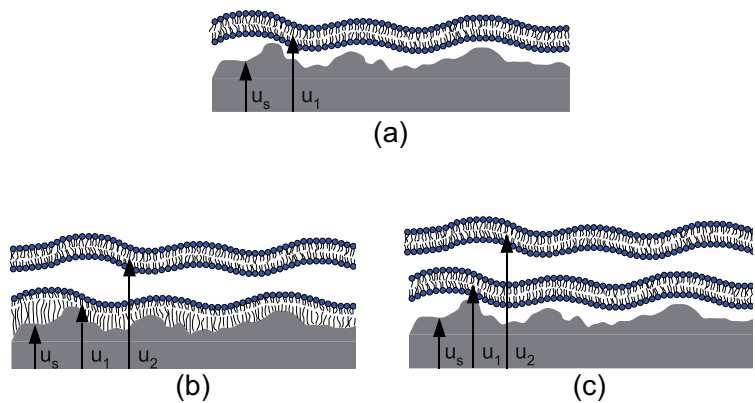}
\caption{Schematic view of supported bilayer systems on a rough substrate: 
(a) single supported bilayer; (b) mixed OTS-lipid supported bilayer; 
(c) supported double bilayer. We use the Monge representation to describe the membranes positions: 
${\bf r_\parallel} = \left(x,y\right)$ is the lateral coordinate, and $z = u_i\left({\bf r_\parallel}\right)$ 
the position of the i$^{th}$ interface (i=s (substrate), 1 (first layer), 2 (floating bilayer)).}.
\label{figsystemes}
\end{center}
\end{figure}
We consider a stack of almost flat membranes, with bending modulus $\kappa_i$ and surface tension $\gamma_i$, 
supported on a rough surface. 
We use the Monge representation to describe the membranes as shown on Fig. \ref{figsystemes}: ${\bf r_\parallel} = \left(x,y\right)$, 
and $z$ is the coordinate perpendicular to the substrate. 
We denote $u_s\left({\bf r_\parallel}\right)$ the substrate position, and $u_i\left({\bf r_\parallel}\right)$ 
the position of the i-th membrane (1 being the closest to the substrate). 
Each membrane interacts with the other components of the system through 
interaction potentials $U_{s,i}\left({\bf r_\parallel}\right)$ for the interaction with the substrate, 
and $U_{i,j}\left({\bf r_\parallel}\right)$ for the interaction with another membrane. 
Following Canham \cite{Canh70} and Helfrich \cite{helfrich73} we write the free energy of the system as:
\begin{eqnarray}
{\cal F}\left[u_i\left( {\bf r_\parallel} \right) \right] = \int d^2{\bf r_\parallel}  \left[  \sum_{i=1}^2 \left( \frac 12 \left( \gamma_i {\bf \nabla}^2 + \kappa_i \Delta^2 \right) u_i\left({\bf r_\parallel}\right) + U_{s,i} \left( {\bf r_\parallel} \right) \right) + U_{1,2} \left( {\bf r_\parallel} \right) \right].
\label{freeenergy}
\end{eqnarray}
Considering small fluctuations of supported bilayers at equilibrium close to a substrate, 
we use a quadratic approximation for the interaction potentials:
\begin{eqnarray}
\nonumber
U_{s,i}\left({\bf r_\parallel}\right) &=& U_{s,i}\left(u_i\left({\bf r_\parallel}\right)-u_s\left({\bf r_\parallel}\right) \right) \simeq A_i \left(u_i\left({\bf r_\parallel}\right)-u_s\left({\bf r_\parallel}\right) \right)^2 \\
U_{i,j}\left({\bf r_\parallel}\right) &=& U_{i,j}\left(u_i\left({\bf r_\parallel}\right)-u_j\left({\bf r_\parallel}\right) \right) \simeq  B \left(u_i\left({\bf r_\parallel}\right)-u_j\left({\bf r_\parallel}\right) \right)^2.
\label{potentiel_quadratique}
\end{eqnarray}
The various contributions to the interaction potential will be discussed in the next section.
This approximation is no longer valid in the regime of large membrane fluctuations (see \cite{lipoleible1}). 
As usual for a system with harmonic coupling between degrees of freedom, 
it is worth rewriting the free energy of the system in Fourier space using:
\begin{eqnarray}
u_i(r_\parallel) = \sum_{q_\parallel} \tilde{u_i}(q_\parallel) e^{i {\bf q_\parallel}.{\bf r_\parallel}},
\label{serie_fourier}
\end{eqnarray}
where ${\bf q_\parallel} =(q_x, q_y)$. 
This transformation allows decoupling of the modes in the Fourier space leading to 
${\cal F} = \sum_{q_\parallel} {\cal F}_{q_\parallel}$ with:
\begin{eqnarray}
{\cal F}_{q_\parallel} &=& \frac 12 \sum_{i=1}^2 \left(\tilde{a}_i(q_\parallel) + B \right) |\tilde{u}_i(q_\parallel)|^2  - A_1 \tilde{u}_1(q_\parallel) \tilde{u}_s(-q_\parallel)- A_2 \tilde{u}_2(q_\parallel) \tilde{u}_s(-q_\parallel) - B \tilde{u}_1(q_\parallel) \tilde{u}_2(-q_\parallel),
\label{freeenergyfourier}
\end{eqnarray}
where we use 
$\tilde{a}_i(q_\parallel) = \left( A_i + \gamma_i q_\parallel^2 + \kappa_i q_\parallel^4 \right)$ with $i=1,2$. 
In the following, this formalism will be used to describe static and thermal deformations of membranes in 
different cases.

\subsection{Interaction potential}
\label{interactionpot}
The physical properties of lipid bilayers are the result of a competition between attractive and repulsive 
molecular interactions (van der Waals, electrostatic, hydration). 
The membranes are also subject to thermal fluctuations which lead to entropic repulsion 
\cite{katsaras,Lipowsky95,israelachvili}.
The van der Waals interaction $U_{vdW}(z)$ between two membranes of thickness $\delta$ separated by a distance 
$z$ can be written:
\begin{eqnarray}
U_{VdW}\left(z\right) = - \frac{H}{12 \pi} \left( \frac{1}{z^2} - \frac{2}{\left(z+\delta\right)^2} + \frac{1}{\left(z+2\delta\right)^2} \right),
\end{eqnarray}
where the Hamaker constant $H$ is on the order of $k_B T$ and depends both on the lipids and on the solvent. 
At short length scales (less than $1$ nm), bilayers separated by a distance $d$ experience 
an exponentially decaying repulsive hydration force, the microscopic origin of which has been the matter
of intense debate \cite{Marcelja1976129,parsegian(lang1991),israelJPhysChem92,besseling(langmuir1997)}.
We write the hydration potential
\begin{eqnarray}
U_{hyd} = P_h d_h exp\left(-\frac{z}{d_h}\right),
\end{eqnarray}
where $d_h \simeq 0.3$ nm \cite{rand(bba1989)} is the hydration length and 
$4\cdot 10^7$ Pa $< P_h < 4 \cdot 10^9$ Pa is the hydration pressure.\\
Finally, as first highligthed by Helfrich \cite{helfrich78}, an additionnal entropic contribution due to the confinment 
of the fluctuating membranes needs to be taken into account. 
Helfrich described the case of purely steric interactions \cite{helfrich78}, leading to the well-known expression: 
\begin{eqnarray}
V_{Helf} \sim \frac{\left(k_B T\right)^2}{\kappa z^2}.
\end{eqnarray}
This expression was extended in a phenomelogical way to tense bilayers \cite{seifert(prl1995)}. 
Podgornik and Parsegian proposed a extension of this theory to soft-confinement, including direct interbilayer interactions \cite{Podgornik1992}.

\section{Correlation functions}
In this section we calculate the height-height auto and cross-correlation functions of the membranes. 
These correlation functions will be used in  Sect. \ref{xray} to calculate the x-ray scattering cross-sections.  
As first described by Swain and Andelman \cite{andelman1999,andelman2001}, a supported bilayer deposited 
on a rough substrate adapts its equilibrium shape to the substrate roughness, and for finite temperatures, 
thermal fluctuations come into play.
We will start by describing the substrate static roughness before considering different
experimental systems (Fig. \ref{figsystemes}): 
single supported bilayer, supported bilayer on a mixed OTS-lipid layer and supported double bilayer. 
In all cases, we first give the static correlation functions, generalizing the Swain and Andelman approach, 
and then describe the thermal contribution to the correlation functions.

\subsection{Static correlation functions}
\subsubsection{Silicon substrate}
\label{substrate}
Like a wide range of rough surfaces, our silicon substrates 
 can be described using self-affine correlation functions \cite{Sinha88}:
\begin{eqnarray}
\left< u_s(0) u_s(r_\parallel) \right> = \sigma_{s}^2 e^{-\left(\frac{r_\parallel}{\xi_{s}}\right)^{2H_{s}}}.
\label{Palazantsas}
\end{eqnarray}
$\sigma_{s}$, is the roughness amplitude, $0 < H_s < 1$ is the roughness exponent which describes the overall shape
of the correlation function and $\xi_s$ is the characteristic length of the roughness (for example, distance between scratches...). 
For distances smaller than $\xi_s$, the correlation function goes to $\sigma_{s}^2$ as 
$\sigma_s^2 (1-(r_\parallel/\xi_s)^{2H_s})$ and decays to $0$ as a stretched or compressed exponential 
for distances larger than $\xi_s$.

\subsubsection{Single supported bilayer}
\label{a membrane near a wall}
We first consider an almost flat single bilayer, with a bending modulus $\kappa_1$ and a surface tension $\gamma_1$, 
interacting with a rough surface (Fig. \ref{figsystemes}(a)). 
The free energy is given by Eq. (\ref{freeenergyfourier}) with N=1. 
To determine the equilibrium state, this equation has to be minimized against $\tilde{u}_1(q)$, 
yielding $\tilde{u}_1(q) = \left( A_1/\tilde{a}_1(q) \right) \tilde{u}_s(q)$. 
The roughness spectrum is given by \cite{andelman1999,andelman2001}:
\begin{eqnarray}
\label{auto-correlation bilayer fourier}
\left< \tilde{u}_1(q_\parallel)\tilde{u}_1(-q_\parallel) \right> &=& \frac{A_1^2}{\tilde{a}_1^2(q_\parallel)} \langle |\tilde{u}_s(q_\parallel)|^2\rangle 
= \frac{A_1^2}{\left( A_1 + \gamma_1 q_\parallel^2 + \kappa_1 q_\parallel^4 \right)^2} \langle |\tilde{u}_s(q_\parallel)|^2\rangle \\
\label{cross-correlation bilayer fourier}
\left< \tilde{u}_1(q_\parallel)\tilde{u}_s(-q_\parallel) \right> &=& \frac{A_1}{\tilde{a}_1(q_\parallel)} \langle |\tilde{u}_s(q_\parallel)|^2\rangle
= \frac{A_1}{A_1 + \gamma_1 q_\parallel^2 + \kappa_1 q_\parallel^4} \langle |\tilde{u}_s(q_\parallel)|^2\rangle.
\end{eqnarray}
For low bending rigidity and/or tension, the membrane is soft enough to follow the substrate roughness and we have 
$\langle |\tilde{u}_1(q_\parallel)|^2\rangle \simeq \langle |\tilde{u}_s(q_\parallel)|^2\rangle$. On the opposite, if the rigidity or the 
membrane tension are large, $\langle |\tilde{u}_1(q_\parallel)|^2\rangle \rightarrow 0$ as the membrane is too stretched 
or rigid to follow the substrate. Two cases can be considered. If $\Delta = \gamma_1^2 -4 A_1\kappa_1 >0$, 
the membrane follows the substrate for wave-vectors $q_\parallel < \sqrt{A_1/\gamma_1}$, there is then 
a tension dominated regime in $1/q_\parallel^2$ for 
$\sqrt{A_1/\gamma_1} < q_\parallel < \sqrt{\gamma_1/\kappa_1}$, and finally a bending rigidity 
dominated regime in $1/q_\parallel^4$ for $q_\parallel >  \sqrt{\gamma_1/\kappa_1}$. If $\Delta < 0$,
we directly go from the potential dominated regime where the membrane follows the substrate to the 
bending rigidity dominated regime.\\
In the real space, Eqs. (\ref{auto-correlation bilayer fourier}) and (\ref{cross-correlation bilayer fourier}) 
become convolution products:
\begin{eqnarray}
\left< u_1(0) u_1(r_\parallel) \right> &=& g_1(r_\parallel) \otimes \left< u_s(0) u_s(r_\parallel) \right> \\
\left< u_1(0) u_1(r_\parallel) \right> &=& h_1(r_\parallel) \otimes \left< u_s(0) u_s(r_\parallel) \right>,
\label{convolution_bilayer}
\end{eqnarray}
where the susceptibilities $g_1(r_\parallel)$ and $h_1(r_\parallel)$ describe how the membrane self- and cross-correlation 
functions correlate to the substrate heght-height correlation function.
$g_1(r_\parallel)$ is the inverse Fourier transform of $A_1^2/\tilde{a}_1^2(q_\parallel)$ and  
$h_1(r_\parallel)$ the inverse Fourier transform of $A_1/\tilde{a}_1(q_\parallel)$. 
In order to calculate $g_1(r_\parallel)$ and $h_1(r_\parallel)$ we need the roots of $\tilde{a}_1(q_\parallel)$ 
given in appendix \ref{coefstatic1bic} in order to decompose $g_1$ and $h_1$ in partial fractions.
In the following we denote them as $q_{1 \parallel}$ and $q_{2 \parallel}$.
The Fourier transforms of polynomial fractions such as $1/(q_\parallel^2+q_{j \parallel}^2)$ or $1/(q_\parallel^2+q_{j \parallel}^2)^2$ are 
proportional to the modified Bessel functions of the second kind $K_0(q_{j \parallel} r_\parallel)$ and $K_1(q_{j \parallel} r_\parallel)$ 
(see appendix \ref{appfourier}) and we obtain:
\begin{eqnarray}
\label{cross-correlation bilayer real}
g_1(r_\parallel) &=& \frac{1}{2\pi} \frac{A_1^2}{\kappa_1^2} \left[\lambda_1 K_0\left(q_{1 \parallel} r_\parallel\right) + \lambda_2 K_0\left(q_{2 \parallel} r_\parallel\right) + \frac {\eta_1^2}{2} \frac{r_\parallel}{q_{1 \parallel}} K_1\left(q_{1 \parallel} r_\parallel\right) + \frac {\eta_2^2}{2} \frac{r_\parallel}{q_{2 \parallel}} K_1\left(q_{2 \parallel} r_\parallel\right)\right]\\
\label{auto-correlation bilayer real}
h_1(r_\parallel) &=& \frac{1}{2\pi} \frac{A_1}{\kappa_1} \left[\eta_1 K_0\left(q_{1 \parallel} r_\parallel\right) + \eta_2 K_0\left(q_{2 \parallel} r_\parallel\right)\right],
\end{eqnarray}
where the $\eta_i$ and $\lambda_i$ are given in App. \ref{appfourier}.

\subsubsection{Two membranes near a wall}
\label{Two membranes near a wall}
In this section, we treat the case of two supported membranes (N=2, Fig.  \ref{figsystemes}(c)). 
The minimization of the free energy against $\tilde{u}_1(q_\parallel)$ and $\tilde{u}_2(q_\parallel)$ gives:
\begin{eqnarray}
\tilde{u}_1(q_\parallel) &=& \frac{A_1 \tilde{a}_2(q_\parallel) + B \left(A_1+ A_2 \right)}{\tilde{a}_1(q_\parallel)\tilde{a}_2(q) + B \left( \tilde{a}_1(q_\parallel)+\tilde{a}_2(q_\parallel)\right)} \tilde{u}_s(q_\parallel)\\
\tilde{u}_2(q_\parallel) &=& \frac{A_2 \tilde{a}_1(q_\parallel) + B \left(A_1+ A_2 \right)}{\tilde{a}_1(q_\parallel)\tilde{a}_2(q_\parallel) + B \left( \tilde{a}_1(q_\parallel)+\tilde{a}_2(q_\parallel)\right)} \tilde{u}_s(q_\parallel).
\end{eqnarray}
Five correlation functions are now needed to describe the membranes: 
(i) the self height-height correlations of each membrane $\left<u_1({0})u_1({r_\parallel})\right>$ and $\left<u_2({0})u_2({r_\parallel})\right>$; 
(ii) the height-height cross-correlations between the membranes and the substrate $\left<u_s({0})u_1({r_\parallel})\right>$ 
and $\left<u_s({0})u_2({r_\parallel})\right>$; 
(iii) and the cross-correlation between the two membranes $\left<u_1({0})u_2({r_\parallel})\right>$. 
These functions can be expressed as a convolution product of Eq. (\ref{Palazantsas}) with a polynomial fraction in $q_\parallel$. 
Applying the method of the previous section, we first determine the roots $q_{j \parallel}$ of the polynomial equation of degree 8, 
$\tilde{a}_1(q_\parallel)\tilde{a}_2(q_\parallel) + B \left( \tilde{a}_1(q_\parallel)+\tilde{a}_2(q_\parallel)\right) = 0$
and perform the partial fraction decomposition. Like in the previous section, the correlation functions can be expressed as combinations 
of the Bessel functions $K_0$ and $K_1$:
\begin{eqnarray}
\left<u_1(0) u_2(r_\parallel) \right> &=& f_{1,2}\left(r\right) \otimes \left< u_s(0) u_s(r_\parallel) \right>\\
\left<u_i(0) u_i(r_\parallel) \right> &=& g_i\left(r\right) \otimes \left< u_s(0) u_s(r_\parallel) \right> \\
\left<u_i(0) u_s(r_\parallel) \right> &=& h_{i,s}\left(r\right) \otimes \left< u_s(0) u_s(r_\parallel) \right>,
\end{eqnarray}
with:
\begin{eqnarray}
\label{cross-correlation quadri quadri}
f_{1,2}\left(r_\parallel\right) &=& \frac{1}{2\pi} \frac{A_1 A_2}{\kappa_1 \kappa_2} \sum_{j=1}^4 \left[ \tau_j K_0(q_{j \parallel} r_\parallel) + \frac 12 \nu_j r_\parallel \frac{K_1(q_{j \parallel} r_\parallel)}{q_{j \parallel}} \right]\\
\label{auto-correlation quadri}
g_i\left(r_\parallel\right) &=& \frac{1}{2\pi} \frac{A_i^2}{\kappa_i^2} \sum_{j=1}^4 \left[ \lambda_{i,j} K_0(q_{j \parallel} r_\parallel) + \frac 12 \eta_{i,j}^2 r_\parallel \frac{K_1(q_{j \parallel} r_\parallel)}{q_{j \parallel}} \right]\\
\label{cross-correlation quadri substrat}
h_{i,s}\left(r_\parallel\right) &=& \frac{1}{2\pi} \frac{A_i}{\kappa_i} \sum_{j=1}^4 \eta_{i,j} K_0(q_{j \parallel} r_\parallel).
\end{eqnarray}
Eqs. (\ref{auto-correlation quadri}) and (\ref{cross-correlation quadri substrat}) are the generalization of the Eqs. (\ref{cross-correlation bilayer real})
 and (\ref{auto-correlation bilayer real}) for double bilayers (see App. \ref{coefstatic2bic} for the expression of the various coefficients). Whatever the number of membranes, these correlation functions will always have the same 
shape. It is worth noting that for $\kappa_2 = \gamma_2 = A_2 = B = 0$, we correctly recover 
the correlation function of the single bilayer obtained in the previous section.\\
In the specific case of symmetrical bilayers ($\gamma_1=\gamma_2$, $\kappa_1=\kappa_2$ and $A_1 = A_2$), we obtain $q_{3 \parallel}=q_{1 \parallel}$ 
and $q_{4 \parallel}=q_{1 \parallel}$, and the parameters $\lambda_{i,j}$ $\eta_{i,j}$, $\tau_j$ and $\nu_j$ can be easily calculated:
\begin{eqnarray}
\nonumber
\tau_1 &=& \tau_2 = \lambda_{1,1} = \lambda_{2,1} = -\frac{2}{\left(q_{2 \parallel}^2 - q_{1 \parallel}^2\right)^3} \mathrm{ \hspace{0.5cm} and \hspace{0.5cm} } \lambda_{1,2} = \lambda_{2,2} = - \tau_1 \\
\nu_1 &=& \frac{1}{\left(q_{2 \parallel}^2 - q_{1 \parallel}^2\right)^2} \mathrm{ \hspace{0.5cm} and \hspace{0.5cm} } \nu_2 = \eta_{1,1}^2 = \eta_{2,1}^2 = \eta_{1,2}^2 = \eta_{2,2}^2 = \nu_1.
\end{eqnarray}
The remaining terms, $\lambda_{i,3}$, $\lambda_{i,4}$, $\eta_{i,3}$, $\eta_{i,4}$, $\nu_3$, $\nu_4$, $\tau_3$ and $\tau_4$ are all equal to zero. 

\subsection{Thermal fluctuations}
In addition to the static roughness induced by the substrate described in the previous section, 
the supported membranes undergo thermal fluctuations which we discuss in this section.
The total displacement of a membrane is  
$\tilde{u}_i(q_\parallel) = \tilde{u}_{i,s}(q_\parallel) + \tilde{u}_{i,th}(q_\parallel)$,
where $\tilde{u}_{i,s}(q_\parallel)$ represents the static position 
of the i-th membrane, and $\tilde{u}_{i,th}(q_\parallel)$ its thermal fluctuations. 
Assuming that there is no correlation between $\tilde{u}_{i,s}(q_\parallel)$ and $\tilde{u}_{j,th}(q_\parallel)$
 we can separate the free energy into a static part ${\cal F}_{s,q_\parallel}$ and a thermal one ${\cal F}_{th,q_\parallel}$.

\subsubsection{Single bilayer}

For a single bilayer, the thermal part of the free energy is ${\cal F}_{th,q_\parallel} = \frac 12 \tilde{a}_{1}(q_\parallel) |\tilde{u}_{th,1}(q_\parallel)|^2$. 
By applying the equipartition theorem one obtains:
\begin{eqnarray}
\left< \tilde{u}_{th,1}(q_\parallel)\tilde{u}_{th,1}(-q_\parallel) \right> = \frac{k_B T}{\tilde{a}_1(q_\parallel)}
= \frac{k_BT}{A_1 + \gamma_1 q_\parallel^2 + \kappa_1 q_\parallel^4} .
\end{eqnarray}
In the real space, the height-height correlation function is simply given by \cite{daillant2005}:
\begin{eqnarray}
\left< u_{th,1}(0) u_{th,1}(r_\parallel) \right> &=& \frac{1}{2\pi} \frac{k_B T}{\kappa_1} \left[ \alpha_{1} K_0(q_{1 \parallel} r_\parallel) +\alpha_{2} K_0(q_{2 \parallel} r_\parallel)\right].
\end{eqnarray}
Coefficients $\alpha_1$ and $\alpha_2$ are given in  App. \ref{thermocoef1bic}.

In the case of a mixed OTS-lipid double bilayer, 
we only have to replace $\tilde{a}_{1}(q_\parallel)$ with $\tilde{b}(q_\parallel)$ in ${\cal F}_{th,q_\parallel}$.

\subsubsection{Double bilayers}

The free energy Eq.(\ref{freeenergy}) for a double bilayer can be written:
\begin{equation}
{\cal F}_{th,q_\parallel} = \frac 12 \sum_{i=1}^2 \left(\tilde{a}_i(q_\parallel) + B \right) |\tilde{u}_{th,i}(q_\parallel)|^2  - B \tilde{u}_{th,1}(q_\parallel) \tilde{u}_{th,2}(-q_\parallel).
\label{freetherm}
\end{equation}
The fluctuation modes of the two membranes $\tilde{u}_{th,1}(q_\parallel)$ and $\tilde{u}_{th,2}(q_\parallel)$ 
are coupled in Eq. (\ref{freetherm}) and the Hamiltonian of the system, 
$$\left(\begin{array}{cc} \tilde{a}_{1}(q_\parallel) + B & -B/2 \\ -B/2 & \tilde{a}_{2}(q_\parallel)+B \end{array}\right)$$
needs first to be diagonalized in order to apply the equipartition theorem. 
The calculation is detailed in App. \ref{appaigen} where we find:
\begin{eqnarray}
\label{cortherm4}
\left< |\tilde{u}_{th,1}(q_\parallel)|^2  \right> &=& k_B T \frac{\tilde{a}_{2}(q_\parallel) + B}{ \tilde{a}_1(q_\parallel) \tilde{a}_2(q_\parallel) + B \left( \tilde{a}_1(q_\parallel)+\tilde{a}_2(q_\parallel)\right)} \nonumber \\
\left< |\tilde{u}_{th,2}(q_\parallel)|^2  \right> & =& k_B T \frac{\tilde{a}_{1}(q_\parallel) + B}{ \tilde{a}_1(q_\parallel) \tilde{a}_2(q_\parallel) + B \left( \tilde{a}_1(q_\parallel)+\tilde{a}_2(q_\parallel)\right)} \\ 
\left< \tilde{u}_{th,1}(q_\parallel) \cdot \tilde{u}_{th,2}(-q_\parallel) \right> &=& k_B T \frac{B}{ \tilde{a}_1(q_\parallel) \tilde{a}_2(q_\parallel) + B \left( \tilde{a}_1(q_\parallel)+\tilde{a}_2(q_\parallel)\right)}. \nonumber
\end{eqnarray}
To determine the inverse Fourier transform of Eqs. (\ref{cortherm4}), 
one first needs to determine the roots $\beta_{i,j}$ of the equation of degree 4, $\tilde{a}_i(q_\parallel) + B = 0$, 
and the roots $q_{i \parallel}$ of the equation of degree 8, 
$\tilde{a}_{1}(q_\parallel) \tilde{a}_{2}(q_\parallel) + B \left( \tilde{a}_1(q_\parallel)+\tilde{a}_2(q_\parallel)\right) = 0$
to perform the partial fraction decomposition. 
It is worth noting that the equation of degree 8 is the same as in the static case. 
Like before, the correlation functions then can be expressed as a combination of the $K_0$ and $K_1$ modified Bessel functions:
\begin{eqnarray}
\left< u_{th,i}(0) u_{th,i}(r_\parallel) \right> &=& \frac{k_B T}{\kappa_i} \frac{1}{2 \pi} \sum_{j=1}^4 \alpha_{i,j} K_0(q_{j \parallel} r_\parallel) \nonumber \\
\left< u_{th,1}(0) u_{th,2}(r_\parallel) \right> &=& \frac{k_B T}{\kappa_1 \kappa_2} \frac{B}{2 \pi} \sum_{j=1}^4 \iota_{j} K_0(q_{j \parallel} r_\parallel),
\label{fctcorre}
\end{eqnarray}
where the parameters are given in App. \ref{thermocoef2bic}.\\
In the particular case of uncoupled bilayers ($B=0$), these equations can be simplified, leading to:
\begin{eqnarray}
\nonumber
q_{1 \parallel} = q_{3 \parallel} = \beta_{1,1} = \beta_{2,1}; \hspace{0.5cm} q_{2 \parallel} = q_{4 \parallel} = \beta_{1,2} = \beta_{2,2}\\
\nonumber
\alpha_{1,1} = \alpha_{2,1} = \frac{1}{q_{2 \parallel}^2 - q_{1 \parallel}^2}; \hspace{0.5cm} \alpha_{1,2} = \alpha_{2,2} = - \alpha_{1,1}.
\end{eqnarray}
All the remaining parameters $\alpha_{i,3}$, $\alpha_{i,4}$ and $\iota_i$ are equal to zero and we recover the results of \cite{daillant2005}:
\begin{eqnarray}
\nonumber
\left< u_{th,i}(0) u_{th,i}(r_\parallel)  \right> &=& \frac{k_B T}{\kappa} \frac{1}{2 \pi} \frac{1}{q_{2 \parallel}^2-q_{1 \parallel}^2} \left( K_0(q_{2 \parallel} r_\parallel) - K_0(q_{1 \parallel} r_\parallel) \right)\\
\nonumber
\left< u_{th,1}(0) u_{th,2}(r_\parallel)  \right> &=& 0.
\end{eqnarray}

\subsection{Discussion}

\begin{figure}[h]
	\begin{minipage}[b]{0.5\linewidth}
		\centering \includegraphics[width=8cm]{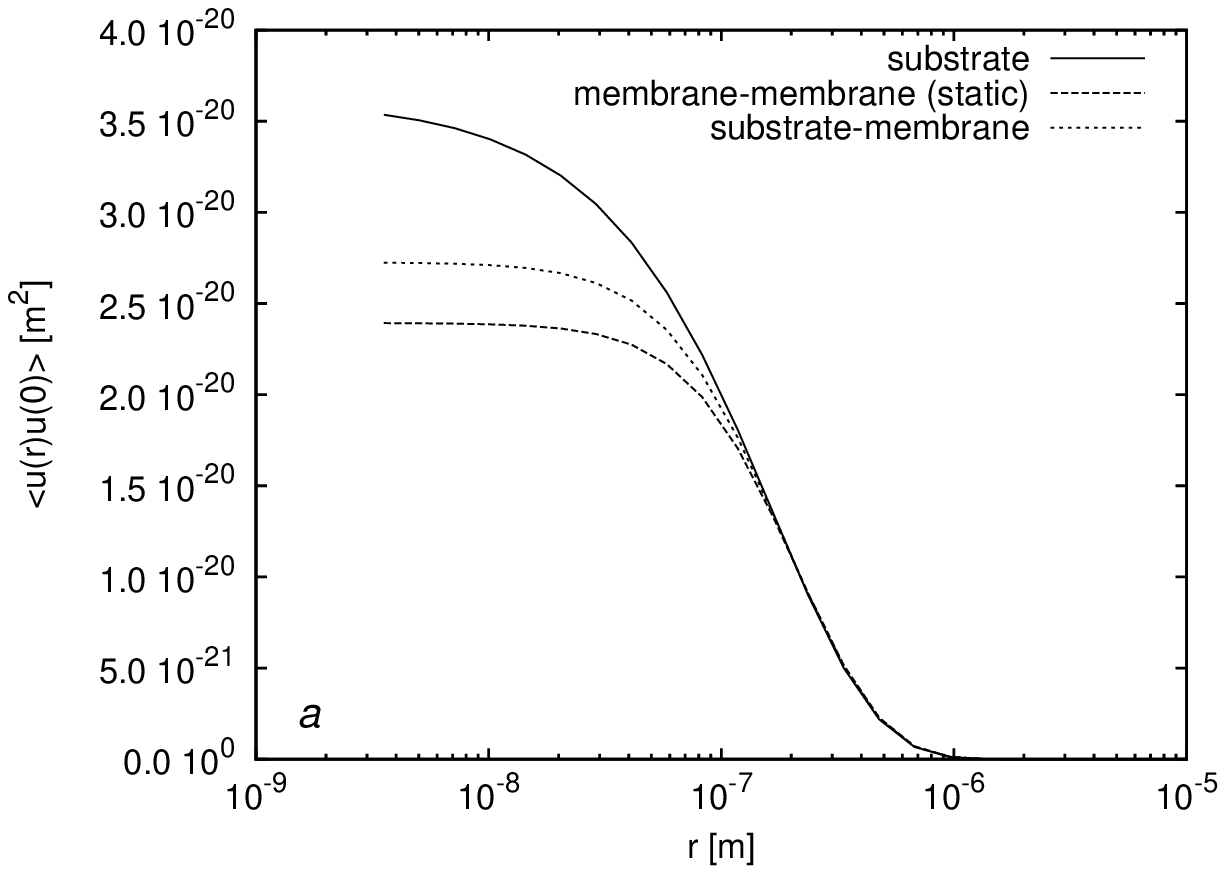}\\
		\centering \includegraphics[width=8cm]{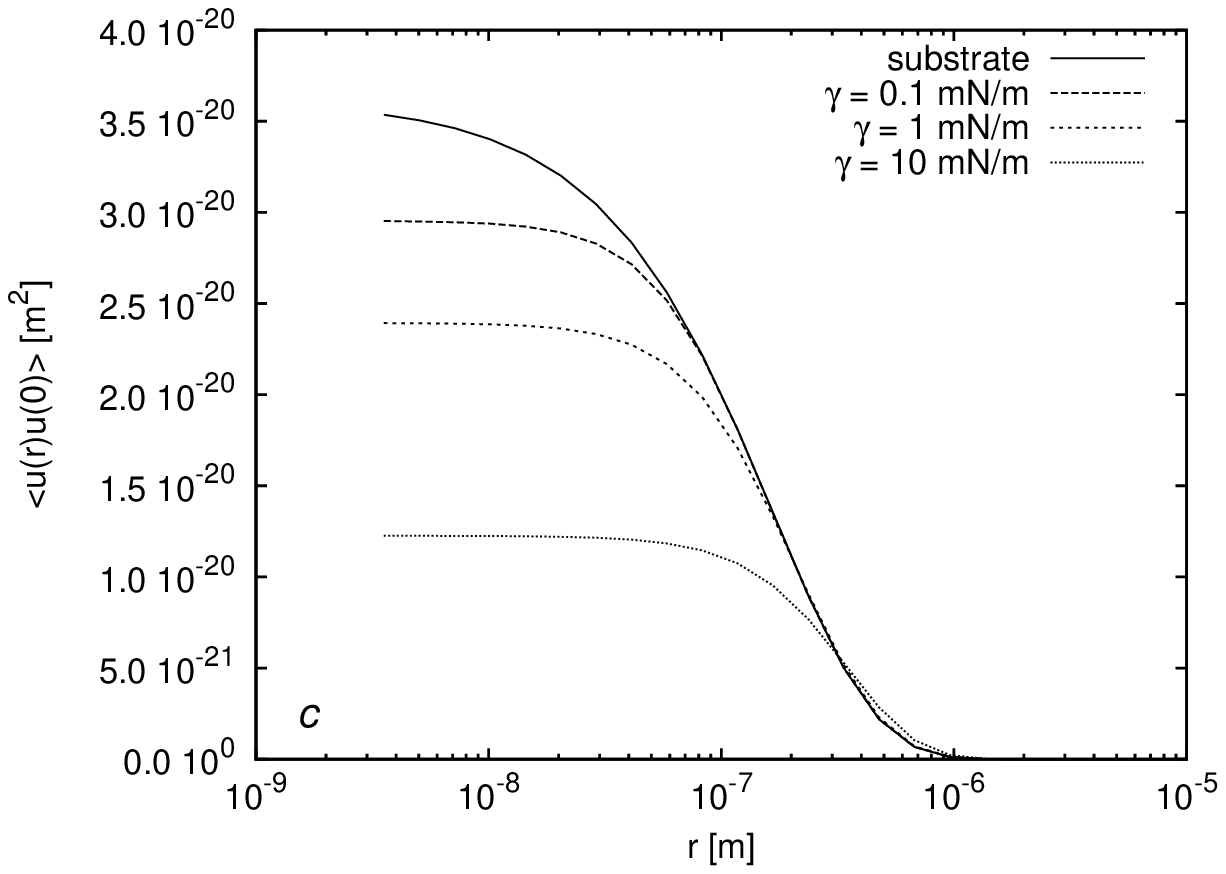}\\
	\end{minipage}\hfill
	\begin{minipage}[b]{0.5\linewidth}
		\centering \includegraphics[width=8cm]{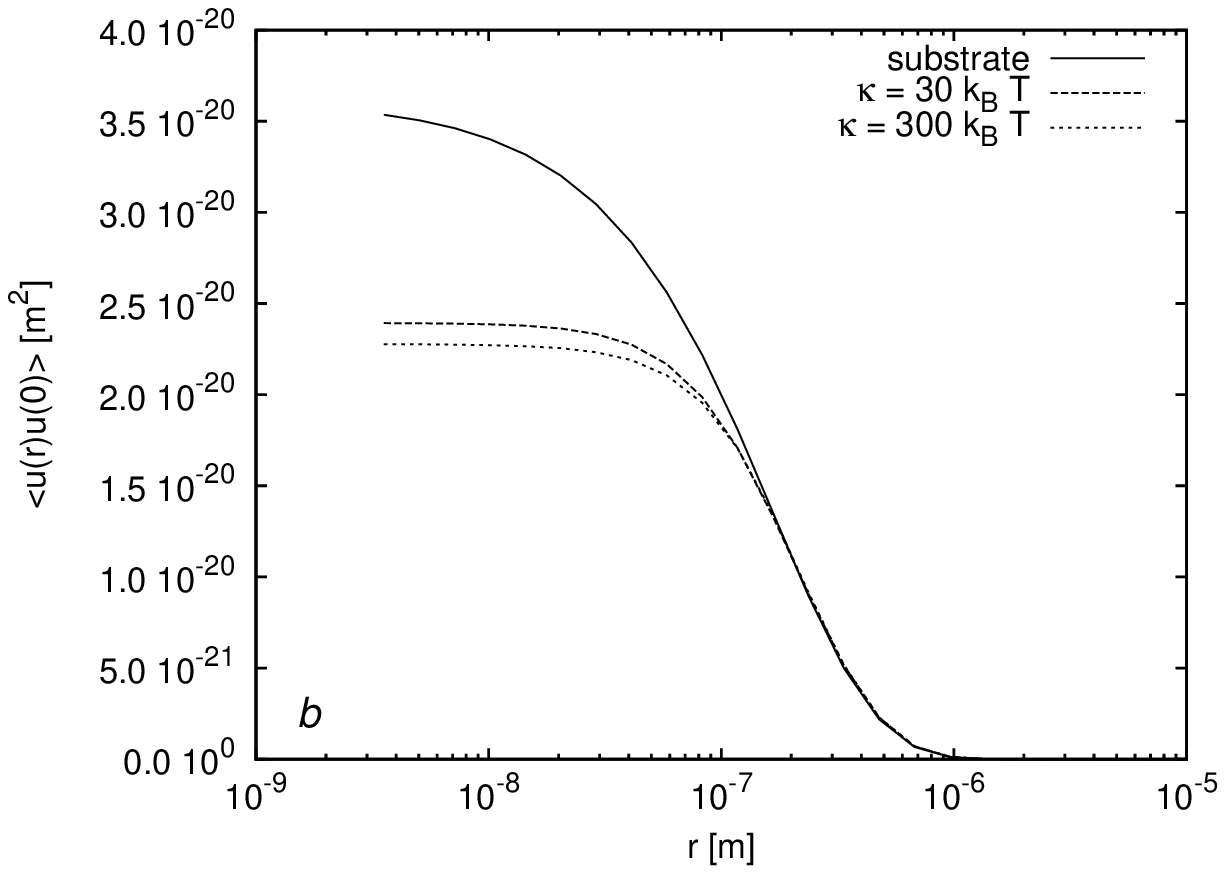}\\
		\centering \includegraphics[width=8cm]{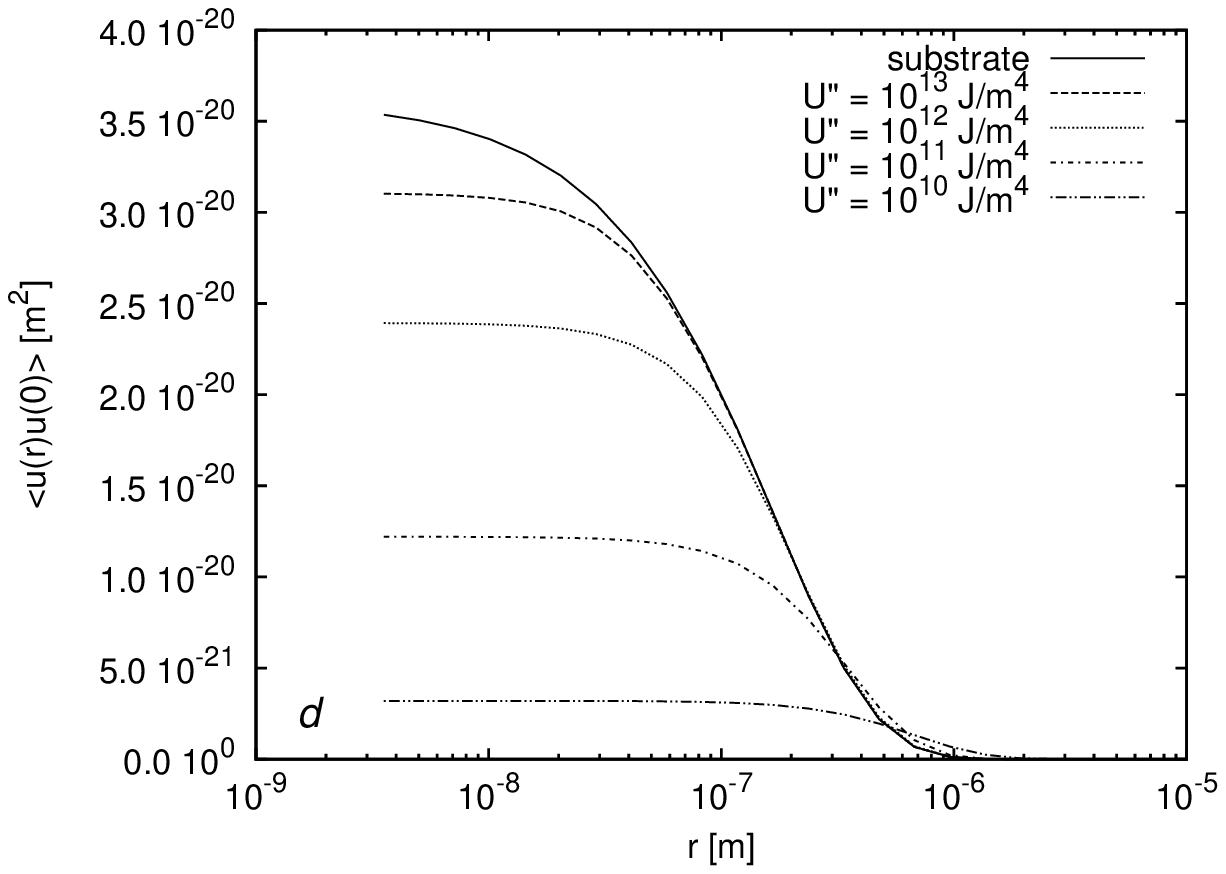}\\
	\end{minipage}
\caption{(a) Static correlation functions for a bilayer with $\gamma = 1$ mN.m$^{-1}$, $\kappa = 30 $ k$_B$T, $A_1 = 10^{12}$ J.m$^{-4}$
on a substrate with $\sigma_s = 0.19$ nm, $\xi_s= 170$ nm and $H_s=0.5$. 
Influence of the different parameters on the bilayer auto-correlation function:  (b) bending rigidity $\kappa_1$,
(c) membrane tension $\gamma_1$, (d) second derivative of the effective potential $A_1$.}
\label{graphes correlation statique}
\end{figure}
Fig. \ref{graphes correlation statique} shows the effect of the elastic properties of a single supported bilayer 
on its coupling to a rough substrate (``static" correlation function). 
The substrate-substrate, substrate-membrane and membrane-membrane 
correlation functions for the bilayer are plotted on Fig.\ref{graphes correlation statique}a. 
In our model, the correlation functions of the membrane (auto-correlation and cross-correlation) are always smaller than that of the substrate and  
the cross-correlation function is always comprised in between the substrate and membrane correlation functions, 
whatever the elastic parameters.  In all cases, the bilayer follows the substrate at 
large lengthscales, whereas the short lengthscale behavior depends on the potential and the elastic properties.\\
Fig. \ref{graphes correlation statique}b shows the membrane-membrane correlation function for two relevant values of the bending rigidity,
$\kappa = 30$ k$_B$T (fluid phase) and $\kappa = 300$ k$_B$T (gel phase). 
The r.m.s. membrane roughness (limit of the correlation function when $r_\parallel \to 0$) decreases for increasing $\kappa$ 
as a more rigid membrane is less free to follow the substrate roughness. 
For the same reason, the cutoff in the susceptibilities $g_1(r_\parallel)$ and $h_1(r_\parallel)$, Eq. (\ref{convolution_bilayer}),
increases with $\kappa$ (from 70 nm for $\kappa = 30$ k$_B$T, $\gamma=1$ mN/m and $A = 10^{12}$ J.m$^{-4}$ to 
209 nm for $\kappa$ = 300 k$_B$T). 
As noticed in Sect. \ref{a membrane near a wall}, in between $\kappa = 30$ k$_B$T and $\kappa = 300$ k$_B$T 
we move from a regime where the cutoff $r_c$ is determined by the tension and rigidity 
($r_c = 2 \pi \sqrt{\kappa / \gamma}$ when $\Delta = \gamma^2 - 4 \kappa A$ is positive) to a regime 
where it is determined by the potential and rigidity ($r_c = 2 \pi (\kappa / A)^{1/4}$ when $\Delta < 0$).\\
Similar trends are observed with increasing tension (Fig. \ref{graphes correlation statique}c) 
and potential (Fig. \ref{graphes correlation statique}d). For example, the cutoff is equal to 
22 nm (roughness ) for $\gamma=0.1$ mN/m $\kappa = 30$ k$_B$T, and $A = 10^{12}$ J.m$^{-4}$), 
70 nm (roughness ) for $\gamma=1$ mN/m and 118 nm (roughness ) for $\gamma=10$ mN/m, as a more stretched membrane 
cannot follow the substrate at short lengthscales. 
Here again, the cutoff is determined by potential and rigidity for $\gamma=0.1$ mN/m and by tension 
and rigidity when $\gamma=1$ mN/m or $\gamma=10$ mN/m.\\
\begin{figure}[h]
	\begin{minipage}[b]{0.5\linewidth}
		\centering \includegraphics[width=8cm]{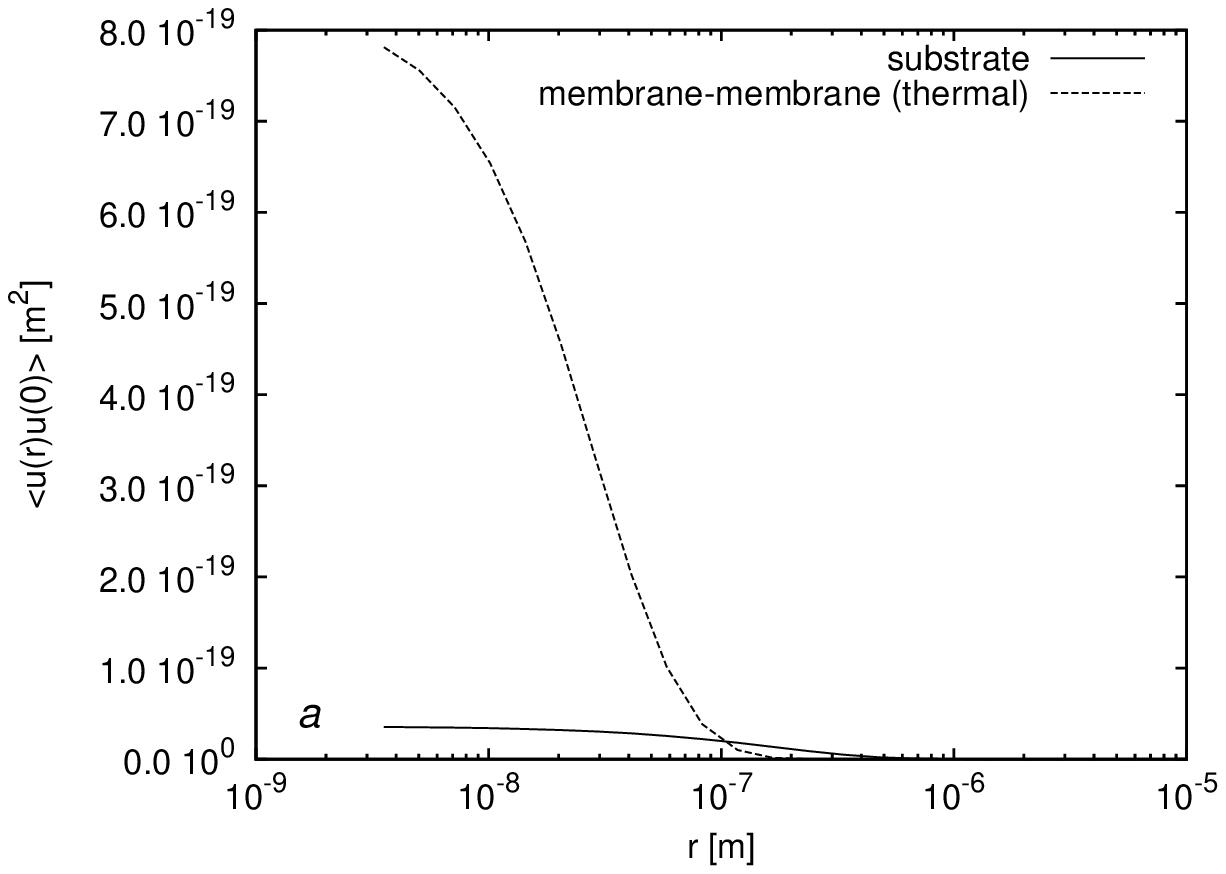}\\
		\centering \includegraphics[width=8cm]{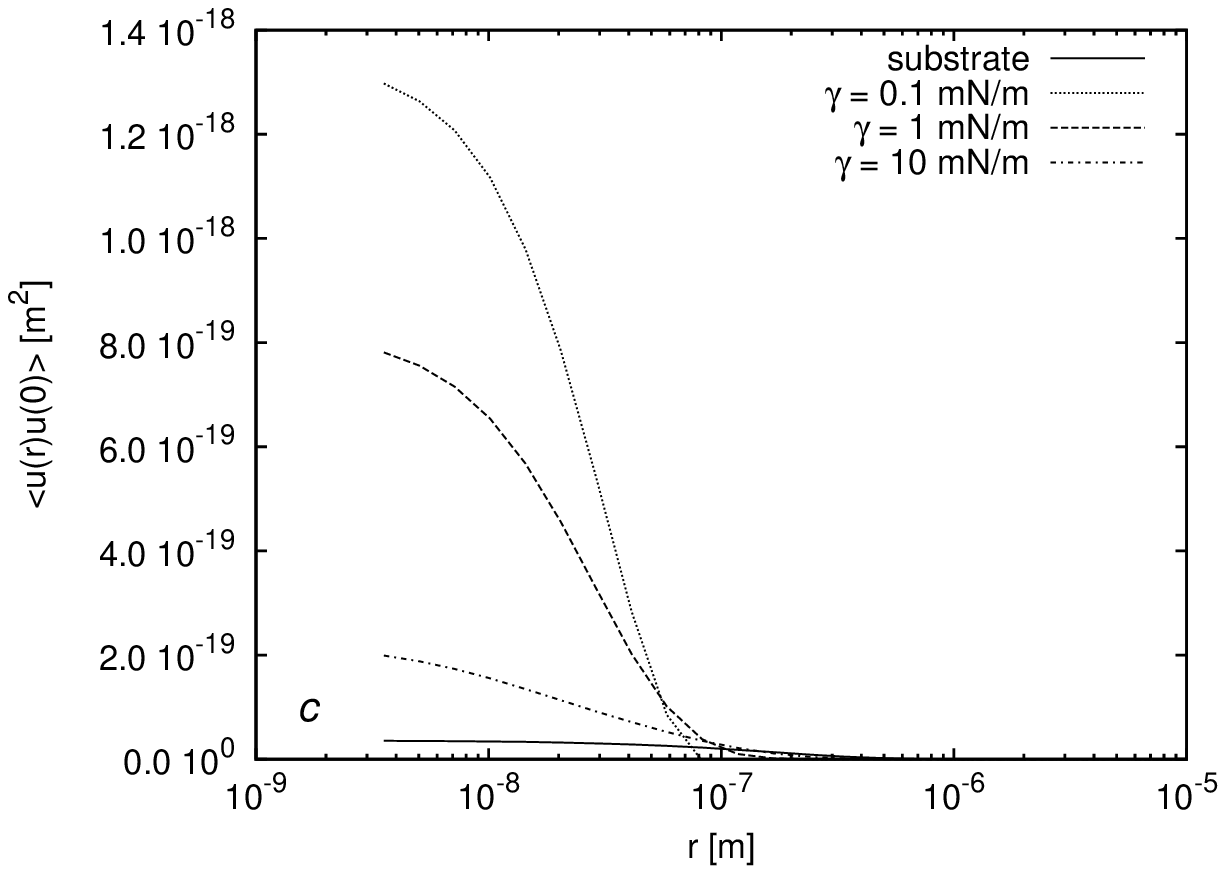}\\
	\end{minipage}\hfill
	\begin{minipage}[b]{0.5\linewidth}
		\centering \includegraphics[width=8cm]{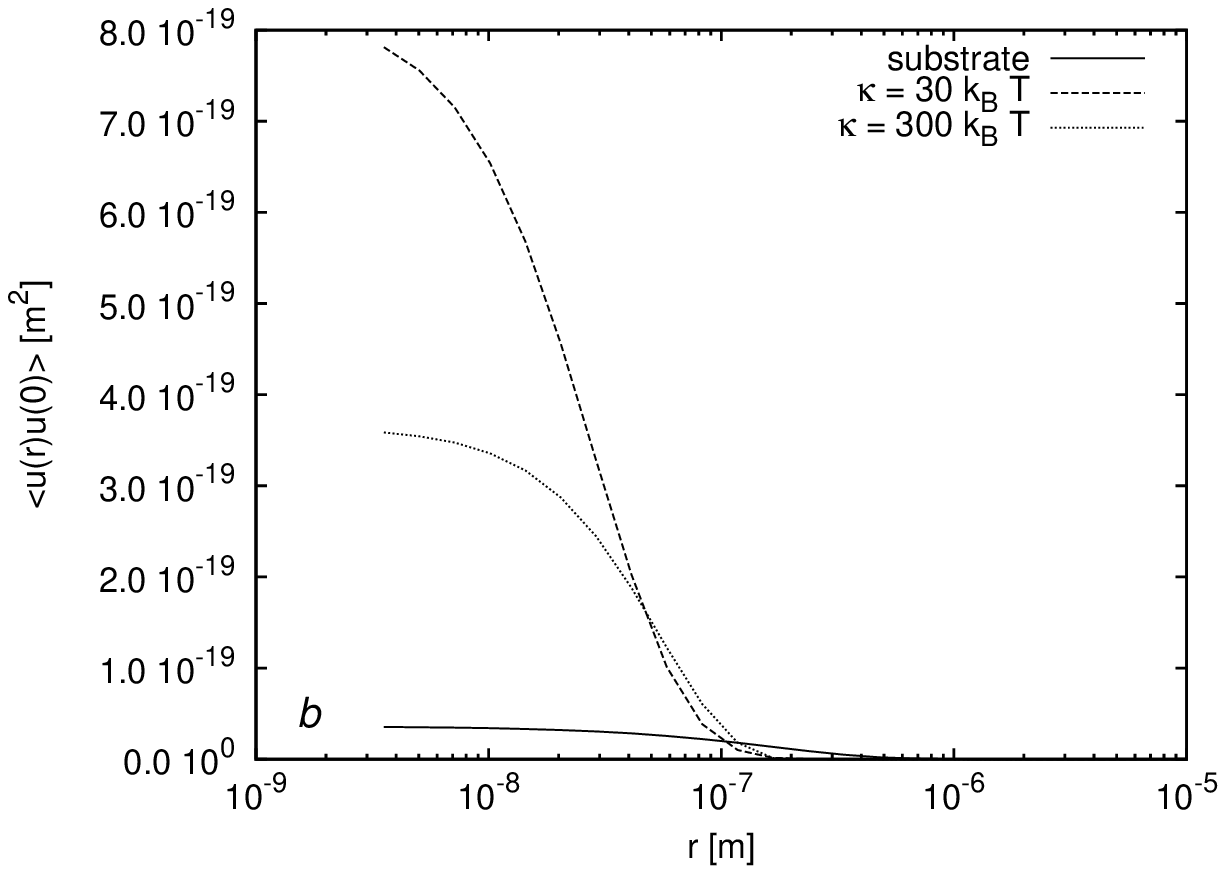}\\
		\centering \includegraphics[width=8cm]{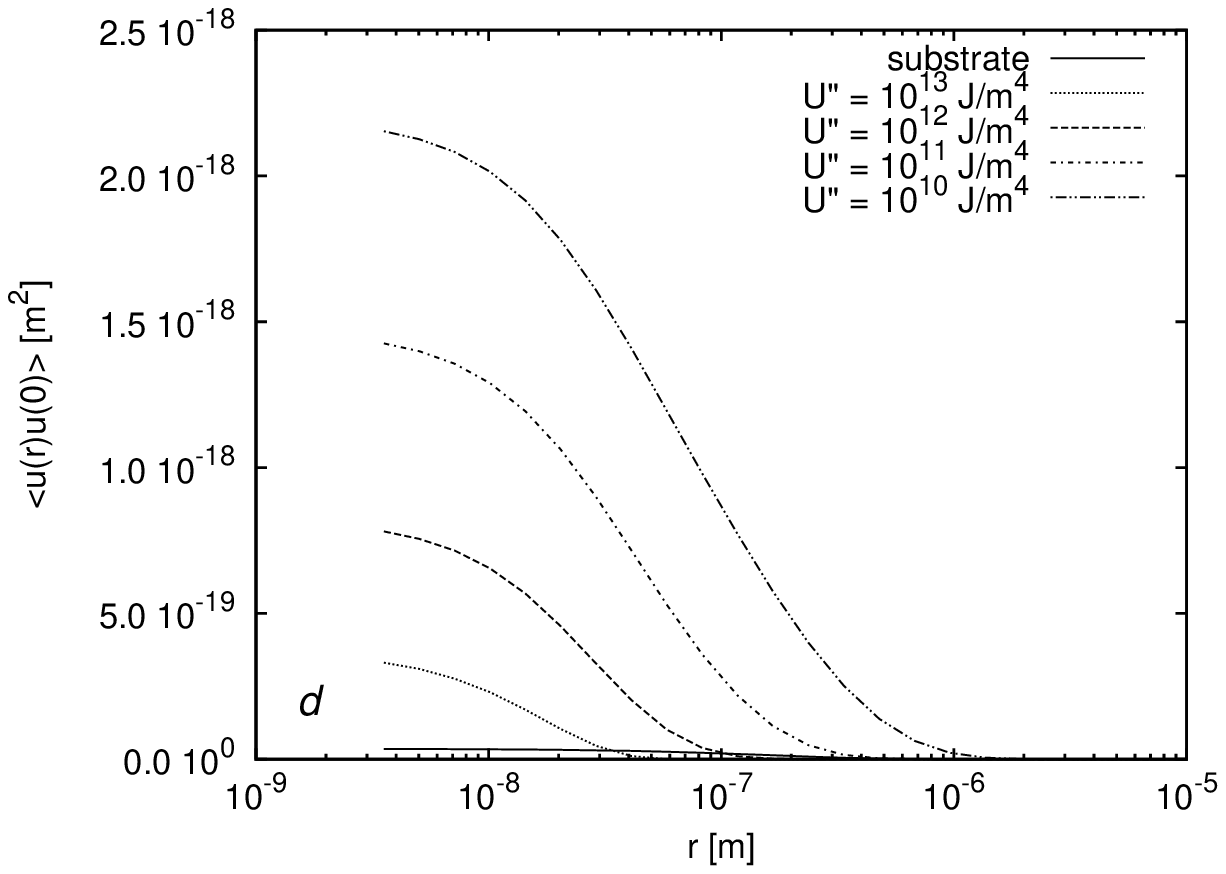}\\
	\end{minipage}
\caption{(a) Thermal correlation functions for a bilayer with $\gamma = 1$ mN.m$^{-1}$, $\kappa = 30 $ k$_B$T, $A_1 = 10^{12}$ J.m$^-4$. 
(b) Effect of the bending rigidity $\kappa$ on the membrane correlation function. (c) Effect of the tension $\gamma$ 
on the membrane correlation function. (d) effect of the second derivative of the effective potential $A_1$ on the 
membrane correlation function.}
\label{graphes correlation thermique}
\end{figure}
The thermal correlation function of a supported bilayer close to a substrate 
is plotted on Fig.\ref{graphes correlation thermique} using the same set of parameters. 
The thermal fluctuations can be larger than the substrate roughness if the interaction potential is soft enough, 
or if the membrane tension and rigidity are small enough.
This is in particular the case in the fluid phase (Fig. \ref{graphes correlation thermique}a), where
the thermal roughness is almost one order of magnitude larger than the static roughness for realistic parameters.
The elastic properties of the membrane have the same effect on static and thermal correlation functions. 
The average roughness of the bilayer and the cutoff increase when the bending rigidity (Fig. \ref{graphes correlation thermique}b),
the tension (Fig. \ref{graphes correlation thermique}c) or the second derivative of the effective potential 
(Fig. \ref{graphes correlation thermique}d) decrease. 
When $\Delta = \gamma^2 - 4 \kappa A >0 $, the membrane tension comes into play, and the correlation function 
has the characteristic logarithmic decay due to its long-range effects.

\section{Specular and off-specular intensity scattered by a membrane}
\label{xray}

\begin{figure}[h]
\begin{minipage}[b]{0.5\linewidth}
\includegraphics[width=6cm]{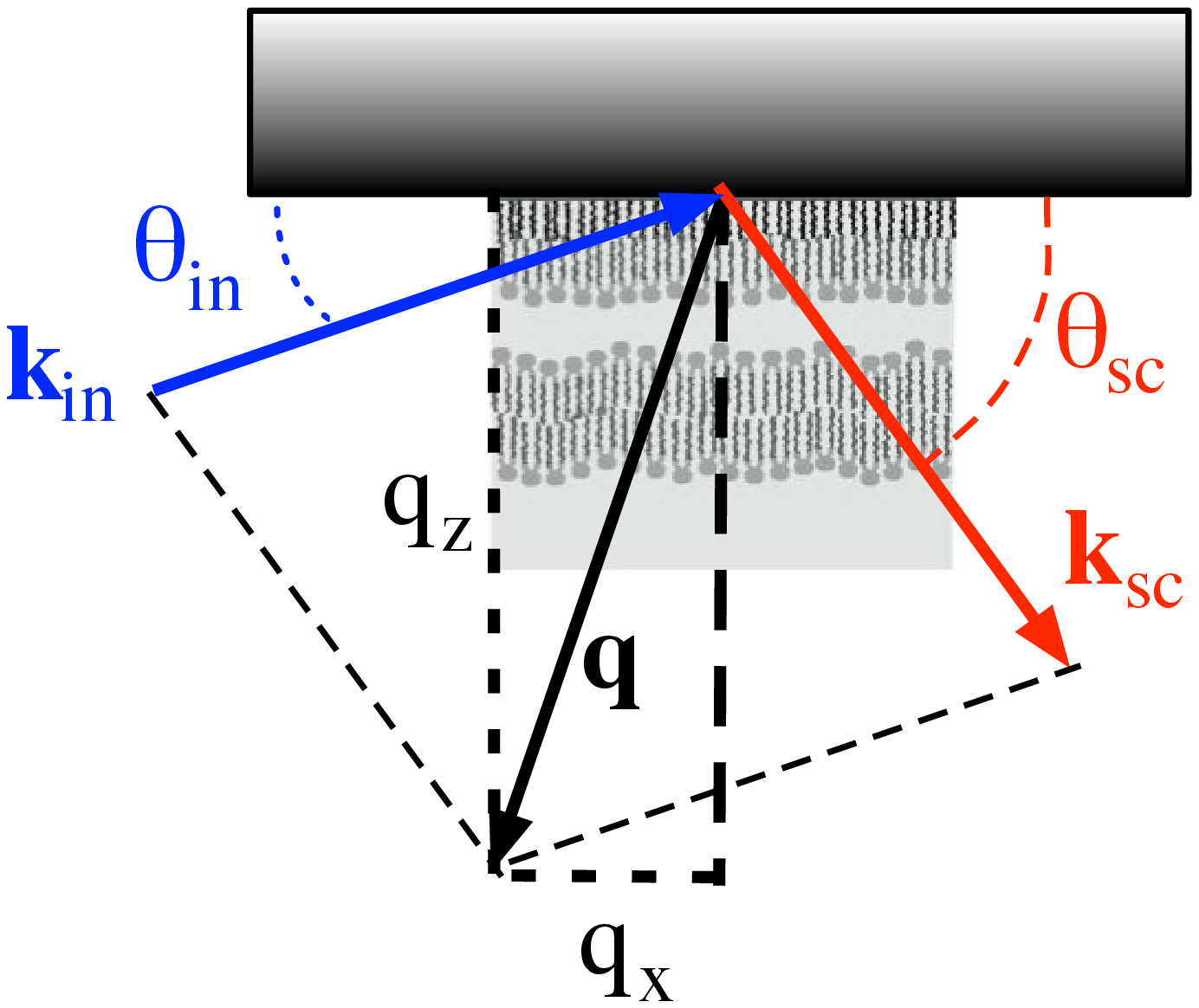}\\ (a) \\
\end{minipage}\hfill
\begin{minipage}[b]{0.5\linewidth}
\includegraphics[width=6cm]{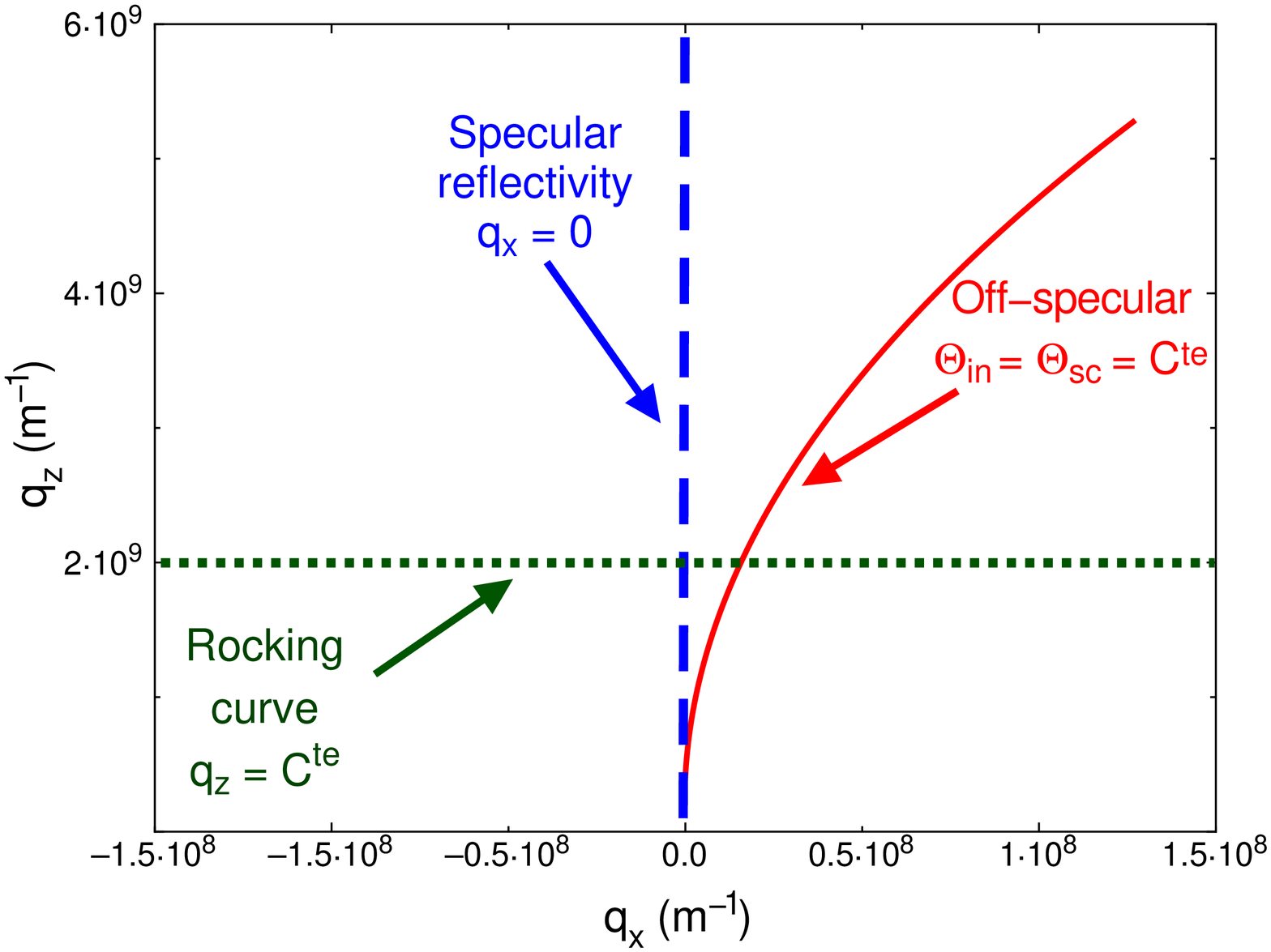}\\ (b)\\
\end{minipage}
\caption{\small{(a) Schematic view of the experimental setup for specular and off-specular reflectivity and 
(b) Fourier space trajectories in specular and off-specular reflectivity experiment. The grazing angle of incidence
is $\theta_{in}$. $\theta_{sc}$ is the angle of scattered x-rays in the plane of incidence ($\psi$ normal to it). }}
\label{setup}
\end{figure}

Specular and off-specular X-ray reflectivity can be used in order to determine the membrane structure and 
properties at submicronic length scales.
Specular reflectivity allows one to determine the structure of the sample perpendicular to the average membrane plane 
whereas off-specular reflectivity also gives access to the elastic properties of the membrane 
(correlation functions, tension, bending energy) and to the interaction potential.\\
In this section, we show how the scattered intensity can be derived by using the correlation functions obtained in the previous section.
The geometry of the experiment is defined on Fig. \ref{setup}. The grazing angle of incidence
is $\theta_{in}$. $\theta_{sc}$ is the angle of the scattered x-rays in the plane of incidence ($\psi$ normal to it).\\
In specular reflectivity experiments, the sample is rocked around the specular condition for every detector position
(thus keeping the normal wave-vector transfer $q_z$ approximately  constant) in order to record and subtract the background.
In off-specular reflectivity, the grazing angle of incidence is kept fixed ($\theta_{in} = 0.7$ mrad) 
below the critical angle for total external reflection $\theta_c \approx 0.85$ mrad for the silicon-water interface at 27 keV, 
as this allows easy background subtraction \cite{daillant2005}, and $\theta_{sc}$ is scanned in the plane of incidence.\\
In order to calculate the intensity scattered by the sample, 
we first calculate the differential scattering cross-section $d\sigma/d\Omega$ which is the power scattered 
by unit solid angle $\Omega_d$ in the direction $\textbf{k}_{sc}$ per unit incident flux in the direction $\textbf{k}_{in}$ (Fig. \ref{setup}a).
The scattered intensity is then calculated by integrating over the detector solid angle,
\begin{eqnarray}
I\left(\textbf{q}\right) = \int d\Omega \frac{d\sigma}{d\Omega}\left(\textbf{q}\right).
\end{eqnarray}
Differential scattering cross-sections are large close to the critical angle for total external reflection and multiple scattering
cannot be neglected. In other words, the simple kinematic Born approximation is no longer valid and a better approximation 
must be used. Here we use a simplified ``Distorted Wave Born Approximation'' (DWBA) \cite{Daillant1999,daillantrop2000, daillant2005}, which 
is a perturbation theory using as reference state a perfectly flat silicon/water interface. As the electron density of lipids 
is close to that of water, this is in fact an excellent approximation.

\subsection{Specular reflectivity}
Within this approximation, the specular reflectivity can be written \cite{Daillant1999,daillant2005}:
\begin{eqnarray}
R\left(q_z\right) = R_F\left(q_z\right) \left|1 + i q_z \int \frac{\delta \rho\left(z\right)}{\rho_{Si} - \rho_{H_2O}} e^{iq_z z} dz\right|^2 = R_F\left(q_z\right) \left|\frac{1}{\rho_{Si} - \rho_{H_2O}} \int 
\left( \frac{\partial \rho}{\partial z} \right)e^{iq_z z} dz\right|^2,
\label{exprefl}
\end{eqnarray}
where $\rho$ is the electron density and $\delta \rho$ is the difference in electron density between the real system 
and the reference state.

\subsection{Off-specular reflectivity}
\subsubsection{Differential scattering cross-section}

Within the DWBA the perturbation part of the differential scattering cross-section is given by \cite{daillant2005}:
\begin{eqnarray}
\left(\frac{d\sigma}{d\Omega}\right) = r_e^2 {\vert t^{in}\vert}^2 {\vert t^{sc}\vert}^2 \left(\textbf{e}_{in}\cdot\textbf{e}_{sc}\right)^2 
\left\langle{\left\vert\int d{\bf r_\parallel} e^{i\textbf{q}\cdot{\bf r_\parallel}} \delta \rho \left(z\right)\right\vert }^2\right\rangle,
\label{cross-section}
\end{eqnarray}
where $r_e= 2.8 \times 10^{-15}$ m 
is the classical radius of the electron, $t^{in} = t_{H_2O,Si}(\theta_{in})$ and $t^{sc} = t_{H_2O,Si}(\theta_{sc})$ 
are the Fresnel transmission coefficients for the silicon/water interface. 
$t^{in}$ is a good approximation to the incident field scattered by the interface, and $t^{sc}$ describes how this field propagates 
to the detector. $\left(\textbf{e}_{in}\cdot\textbf{e}_{sc}\right)^2$ is the polarization factor. 
In our case $\left(\textbf{e}_{in}\cdot\textbf{e}_{sc}\right)^2 \simeq 1$ \cite{daillantrop2000}.
As our sample is composed of a substrate and of two bilayers, we can split the electron density in three terms: 
$\delta \rho_\mathrm{{sub}}$ for the substrate and 
$\delta \rho_{\mathrm{M1}}$ and $\delta \rho_{\mathrm{M2}}$ for the two membranes.
Each of these terms depends on ${\bf r_\parallel}$ because of the static roughness and of the fluctuations. We have 
\begin{eqnarray}
\delta \rho \left(z,{\bf r_\parallel}\right) = \delta \rho_{\mathrm{sub}}\left(z-u_s\right) + \delta \rho_{\mathrm{M1}}\left(z-u_{1,th}-u_{1,st}\right) + \delta \rho_{\mathrm{M2}}\left(z-u_{2,th}-u_{2,st}\right),
\end{eqnarray}
where $u_s$, $u_{i,st}$ and $u_{i,th}$ depend on ${\bf r_\parallel}$.
$\delta \rho_{\mathrm{sub}} = \rho_{\mathrm{sub}} - \rho_{H_2O}$ for $ 0 < z < u_s$  if $u_s >0$ and 
$\delta \rho_{\mathrm{sub}} = \rho_{H_2O}- \rho_{\mathrm{sub}}$ for $ u_s < z < 0$ if $u_s<0$. 
$\delta \rho_{\mathrm{Mi}} = \rho_{\mathrm{Mi}} - \rho_{H_2O}$ with $i=1,2$ are the electron density profiles of membranes relative to water.
After change of variables $z-u_{i,th}-u_{i,st} \to z$, we obtain:
\begin{eqnarray}
\nonumber
\left(\frac{d\sigma}{d\Omega}\right)_{\mathrm{}} &=& r_e^2 {\vert t^{in}\vert}^2 {\vert t^{sc}\vert}^2 \left(\textbf{e}_{in}\cdot\textbf{e}_{sc}\right)^2
\left\langle{\left\vert \int d{\bf r_\parallel} e^{i \textbf{q}_\parallel {\bf r_\parallel}}\left(\delta\tilde{\rho}_{\mathrm{sub}}\left(q_z\right) e^{i q_z u_s\left({\bf r_\parallel}\right)} \right.\right.}\right. \\
&+& \left.{\left.\left. \delta\tilde{\rho}_{\mathrm{M1}}\left(q_z\right) e^{i q_z\left(u_{1,th}\left({\bf r_\parallel}\right)+ u_{1,st}\left({\bf r_\parallel}\right)\right)} + \delta\tilde{\rho}_{\mathrm{M2}}\left(q_z\right) e^{i q_z\left(u_{2,th}\left({\bf r_\parallel}\right)
+ u_{2,st}\left({\bf r_\parallel}\right)\right)} \right) \right\vert}^2 \right\rangle,
\label{eqdsig2}
\end{eqnarray}
where $\delta\tilde{\rho}_{\mathrm{sub}}\left(q_z\right)$, $\delta\tilde{\rho}_{\mathrm{M1}}\left(q_z\right)$ and 
$\delta\tilde{\rho}_{\mathrm{M2}}\left(q_z\right)$ are the substrate and bilayer form factors. 
Assuming that there is no cross correlation between the thermal and static fluctuations of the membranes and between the
thermal fluctuations of the membranes and the substrate roughness, we get:
\begin{eqnarray}
\nonumber
\left\langle  \left[ u_{i,th}\left(\textbf{0}\right) 
+ u_{i,st}\left(\textbf{0}\right)\right]\left[ u_{i,th}\left({\bf r_\parallel}\right) + u_{i,st}\left({\bf r_\parallel}\right)\right]\right\rangle &=& \left\langle u_{i,th}\left(\textbf{0}\right)u_{i,th}\left({\bf r_\parallel}\right) \right\rangle 
+ \left\langle u_{i,st}\left(\textbf{0}\right)u_{i,st}\left({\bf r_\parallel}\right) \right\rangle\\
\left\langle u_s\left(\textbf{0}\right)\left[u_{i,th}\left({\bf r_\parallel}\right) + u_{i,st}\left({\bf r_\parallel}\right)\right] \right\rangle &=& \left\langle u_s\left(\textbf{0}\right)u_{i,st} \left({\bf r_\parallel}\right) \right\rangle.
\end{eqnarray}
We then proceed by developping Eq. (\ref{eqdsig2}). Using the classical result $\left<e^{iqz}\right> = e^{-q^2\left<z^2\right>/2}$ for a 
Gaussian variable z, we can rewrite the differential scattering cross-section as:
\begin{eqnarray}
\nonumber
\left(\frac{d\sigma}{d\Omega}\right) &=& {\cal A} r_e^2 {\vert t^{in}\vert}^2 {\vert t^{sc}\vert}^2 \left(\textbf{e}_{in}\cdot\textbf{e}_{sc}\right)^2 \int d{\bf r_\parallel} e^{i \textbf{q}_\parallel {\bf r_\parallel}} \left[ {\left\vert \delta\tilde{\rho}_{\mathrm{sub}}\left(q_z\right)\right\vert}^2 e^{-q_z^2 \sigma_s^2} e^{q_z^2 \left\langle u_s\left(0\right) u_s\left({\bf r_\parallel}\right) \right\rangle}
\right. \\
\nonumber
&+&\left. 
 {\left\vert \delta\tilde{\rho}_{\mathrm{M1}}\left(q_z\right) \right\vert}^2 e^{-q_z^2 \left( \sigma_{1,th}^2 + \sigma_{1,st}^2\right)} e^{q_z^2 \left\langle u_{1,th} \left(0\right)u_{1,th}\left({\bf r_\parallel}\right) \right\rangle} e^{q_z^2 \left\langle u_{1,st}\left(0\right) u_{1,st}\left({\bf r_\parallel}\right)  \right\rangle}
\right. \\
\nonumber
&+&\left. 
 {\left\vert \delta\tilde{\rho}_{\mathrm{M2}}\left(q_z\right)\right\vert}^2 e^{-q_z^2 \left( \sigma_{2,th}^2 + \sigma_{2,st}^2\right)} e^{q_z^2 \left\langle u_{2,th}\left(0\right) u_{2,th} \left({\bf r_\parallel}\right) \right\rangle} e^{q_z^2 \left\langle u_{2,st}\left(0\right) u_{2,st}\left({\bf r_\parallel}\right) \right\rangle}
\right. \\
\nonumber
&+&\left. 
 \left(\delta\tilde{\rho}_{\mathrm{M1}}^{\ast}\left(q_z\right) \delta\tilde{\rho}_{\mathrm{M2}}\left(q_z\right) + \delta\tilde{\rho}_{\mathrm{M1}}\left(q_z\right) \delta\tilde{\rho}_{\mathrm{M2}}^{\ast}
\left(q_z\right) \right) e^{-\frac 12 q_z^2 \left( \sigma_{1,th}^2 + \sigma_{1,st}^2 + \sigma_{2,th}^2 + \sigma_{2,st}^2\right)} e^{q_z^2 \left\langle  u_{2,th}\left(0\right)u_{1,th}\left({\bf r_\parallel}\right) \right\rangle} e^{q_z^2 \left\langle u_{2,st}\left(0\right)u_{1,st}\left({\bf r_\parallel}\right) \right\rangle}
\right.\\
\nonumber
&+&\left. 
 \left(\delta\tilde{\rho}_{\mathrm{M1}}^{\ast}\left(q_z\right) \delta\tilde{\rho}_{\mathrm{sub}}\left(q_z\right) + \delta\tilde{\rho}_{\mathrm{M1}}
\left(q_z\right) \delta\tilde{\rho}_{\mathrm{sub}}^{\ast}\left(q_z\right) \right) e^{-\frac 12 q_z^2 \left( \sigma_{1,th}^2 + \sigma_{1,st}^2 + \sigma_s^2\right)} 
e^{q_z^2 \left\langle u_s\left(0\right) u_{1,st}\left({\bf r_\parallel}\right) \right\rangle}
\right. \\
\nonumber
&+&\left. 
\left(\delta\tilde{\rho}_{\mathrm{M2}}^{\ast}\left(q_z\right) \delta\tilde{\rho}_{\mathrm{sub}}
\left(q_z\right) + \delta\tilde{\rho}_{\mathrm{M2}}\left(q_z\right) \delta\tilde{\rho}_{\mathrm{sub}}^{\ast}\left(q_z\right) \right) e^{-\frac 12 q_z^2 \left( \sigma_{2,th}^2 + 
\sigma_{2,st}^2 + \sigma_s^2\right)} e^{q_z^2 \left\langle u_s\left(0\right)u_{2,st}\left({\bf r_\parallel}\right) \right\rangle} \right],\\
\end{eqnarray}
where $\sigma_{i,st}^2 = \left<u_{i,st}\left(\textbf{0}\right)^2\right>$  and $\sigma_{i,th}^2 = 
\left<u_{i,th}\left(\textbf{0}\right)^2\right>$ are respectively the static and the thermal surface roughness of the i-th membrane. 
${\cal A}$ is the illuminated area on the sample \cite{thesemora}.
Finally, rewriting $e^{x}=1+e^{x}-1$, specular and off-specular parts of the perturbation in the scattering cross-section can be separated:
\begin{eqnarray}
\left(\frac{d\sigma}{d\Omega}\right) &=& {\cal A} r_e^2 {\vert t^{in}\vert}^2 {\vert t^{sc}\vert}^2 \left(\textbf{e}_{in}\cdot\textbf{e}_{sc}\right)^2 \left[g\left(q_z\right) \left(2 \pi\right)^2 \delta\left(\textbf{q}_\parallel\right) + \int d{\bf r_\parallel} e^{i \textbf{q}_\parallel {\bf r_\parallel}} f\left({\bf r_\parallel}, q_z \right) \right],
\end{eqnarray}
with:
\begin{eqnarray}
\nonumber
f\left({\bf r_\parallel}, q_z \right) &=& {\left\vert\tilde{\rho}_{\mathrm{sub}}\left(q_z\right)\right\vert}^2 e^{-q_z^2 \sigma_s^2} 
\left(e^{q_z^2 \left\langle u_s\left(0\right)u_s\left({\bf r_\parallel}\right) \right\rangle} - 1\right) \\
\nonumber
&+& {\left\vert\tilde{\rho}_{\mathrm{M1}}\left(q_z\right) 
\right\vert}^2 e^{-q_z^2 \left( \sigma_{1,th}^2 + \sigma_{1,st}^2\right)} \left(e^{q_z^2 \left\langle u_{1,th} \left(0\right)u_{1,th}\left({\bf r_\parallel}\right) \right\rangle} e^{q_z^2 \left\langle u_{1,st}\left(0\right)u_{1,st}\left({\bf r_\parallel}\right) \right\rangle}  - 1\right) \\
\nonumber
&+&{\left\vert\tilde{\rho}_{\mathrm{M2}}\left(q_z\right)\right\vert}^2 e^{-q_z^2 \left( \sigma_{2,th}^2 + \sigma_{2,st}^2\right)} \left(e^{q_z^2 \left\langle u_{2,th}\left(0\right)u_{2,th} \left({\bf r_\parallel}\right) \right\rangle} e^{q_z^2 \left\langle u_{2,st}\left(0\right)u_{2,st}\left({\bf r_\parallel}\right) \right\rangle}  - 1\right) \\
\nonumber
&+& \left(\tilde{\rho}_{\mathrm{M1}}^{\ast}\left(q_z\right)\tilde{\rho}_{\mathrm{M2}}\left(q_z\right) + \tilde{\rho}_{\mathrm{M1}}\left(q_z\right)\tilde{\rho}_{\mathrm{M2}}^{\ast}
\left(q_z\right) \right) e^{-\frac 12 q_z^2 \left( \sigma_{1,th}^2 + \sigma_{1,st}^2 + \sigma_{2,th}^2 + \sigma_{2,st}^2\right)} \left(e^{q_z^2 \left\langle u_{2,th}\left(0\right)u_{1,th}\left({\bf r_\parallel}\right) \right\rangle} e^{q_z^2 \left\langle u_{2,st}\left(0\right)u_{1,st}\left({\bf r_\parallel}\right) \right\rangle} - 1\right) \\
\nonumber
&+& \left(\tilde{\rho}_{\mathrm{M1}}^{\ast}\left(q_z\right)\tilde{\rho}_{\mathrm{sub}}\left(q_z\right) + \tilde{\rho}_{\mathrm{M1}}
\left(q_z\right)\tilde{\rho}_{\mathrm{sub}}^{\ast}\left(q_z\right) \right) e^{-\frac 12 q_z^2 \left( \sigma_{1,th}^2 + \sigma_{1,st}^2 + \sigma_s^2\right)} \left(e^{q_z^2 \left\langle u_s\left(0\right)u_{1,st}\left({\bf r_\parallel}\right) \right\rangle} - 1\right) \\
&+& \left(\tilde{\rho}_{\mathrm{M2}}^{\ast}\left(q_z\right)\tilde{\rho}_{\mathrm{sub}}
\left(q_z\right) + \tilde{\rho}_{\mathrm{M2}}\left(q_z\right)\tilde{\rho}_{\mathrm{sub}}^{\ast}\left(q_z\right) \right) e^{-\frac 12 q_z^2 \left( \sigma_{2,th}^2 + 
\sigma_{2,st}^2 + \sigma_s^2\right)} \left(e^{q_z^2 \left\langle  u_s\left(0\right)u_{2,st}\left({\bf r_\parallel}\right) \right\rangle} - 1\right).
\label{fonctionf}
\end{eqnarray}
The $g(q_z)$ function will not be used in the following as we prefer to use Eq. (\ref{exprefl}) which directly gives access to 
the total reflectivity and not only to the perturbation part.

\subsubsection{Off-specular intensity}
Resolution effects can be taken into account by using a resolution function 
${\cal R}es\left(\theta_{sc},\psi \right)$, which is equal to one if $\left(\theta_{sc},\psi \right)$ points to 
the detector area and 0 outside, 
\begin{eqnarray}
I &=& \frac{I_0}{h_i w_i} \int \left[ \left(\frac{d\sigma}{d\Omega}\right)_{\mathrm{spec}}\left(\textbf{q}_\parallel \right) 
+ \left(\frac{d\sigma}{d\Omega}\right)_{\mathrm{off-spec}}\left(\textbf{q}_\parallel \right) \right] 
{\cal R}es\left(\theta_{sc},\psi \right) d\Omega.
\label{intensite_omega}
\end{eqnarray}
$h_i$ and $w_i$ are the incident beam height and width so that $I_0 /h_i w_i$ is the incident flux.
Contrary to the specular case where resolution effects amount to a convolution, the off-specular intensity is proportional 
to the detector solid angle.
The experimental resolution is a unit rectangular function centered on the detector, of width $\Delta\theta$ in the plane 
of incidence and $\Delta\psi$ perpendicular to it. It is in fact easier to proceed in the Fourier space where we have:
\begin{eqnarray}
q_x = k_0 \left( \cos\theta_{in} - \cos \theta_{sc} \right) &\Rightarrow& dq_x = k_0 \sin\theta_{sc} d\theta \\
q_y = k_0 \sin\psi \simeq k_0 \psi &\Rightarrow& dq_y = k_0 d\psi.
\label{chgt_variable}
\end{eqnarray}
For off-specular scattering, Eq. (\ref{intensite_omega}) becomes:
\begin{eqnarray}
\nonumber
I_{\mathrm{off-spec}} &=& \frac{I_0}{h_i w_i k_0^2 \sin \theta_{sc}} \int \left(\frac{d\sigma}{d\Omega}\right)_{\mathrm{off-spec}} \tilde{{\cal R}es}\left(\textbf{q}_\parallel \right) d\textbf{q}_\parallel \\
 &=& \frac{I_0}{h_i w_i} \frac {{\cal A} r_e^2 {\vert t^{in}\vert}^2 {\vert t^{sc}\vert}^2 \left(\textbf{e}_{in}\cdot\textbf{e}_{sc}\right)^2} {k_0^2 \sin\theta_{sc}} \int dx \int dy \int dq_x \int dq_y e^{i q_x x} e^{i q_y y} f\left({\bf r_\parallel}, q_z \right) \tilde{{\cal R}es}\left(\textbf{q}_\parallel\right).
\end{eqnarray}
Since, under our experimental conditions, the slits are widely open in the $y$ direction, 
we have $\tilde{{\cal R}es}(\textbf{q}_\parallel) = \tilde{{\cal R}es}(q_x)$. 
The integration over $q_y$ and $y$ is performed using $\int dq_y e^{i q_y y} = 2 \pi \delta \left(y \right)$. We first get:
\begin{eqnarray}
I_{\mathrm{off-spec}} = \frac{I_0}{h_i w_i} \frac {2 \pi {\cal A} r_e^2 {\vert t^{in}\vert}^2 {\vert t^{sc}\vert}^2 \left(\textbf{e}_{in}\cdot\textbf{e}_{sc}\right)^2} {k_0^2 \sin\theta_{sc}} \int dx \int dq_x e^{i q_x x} f\left(\left( x,0 \right), q_z \right)\tilde{{\cal R}es}\left(q_x\right).
\end{eqnarray}
We further proceed by approximating the resolution function along  $q_x$ as a Gaussian function of width 
$\Delta q_x = k_0 \sin \theta_{sc} \Delta \theta_{sc}$ centered in $q_{x_0}$:
\begin{equation}
\tilde{{\cal R}es}\left(q_x \right) = \frac{1}{\sqrt{2\pi}} e^{-\frac{\left(q_x - q_{x_0}\right)^2}{2 {\Delta q_x}^2}}.
\end{equation}
The term $1/\sqrt{2\pi}$ is used to normalize the function. One obtains:
\begin{equation}
I_{\mathrm{off-spec}} = \frac{I_0}{h_i w_i} \frac {\sqrt{2 \pi} {\cal A} r_e^2 {\vert t^{in}\vert}^2 {\vert t^{sc}\vert}^2 \left(\textbf{e}_{in}\cdot\textbf{e}_{sc} \right)^2} {k_0^2 \sin\theta_{sc}} \int dq_x \int dx e^{i q_{x_0} x} e^{i q_x x} e^{-\frac{q_x^2}{2 {\Delta q_x}^2}} f\left(x, q_z \right).
\end{equation}
Integrating over $q_x$ gives:
\begin{equation}
\int dq_x e^{i q_x x} e^{-\frac {q_x^2} {2 {\Delta q_x}^2}} = \sqrt{2 \pi} \Delta q_x e^{-\frac{{\Delta q_x}^2 x^2}{2}}.
\end{equation}
We finally obtain:
\begin{eqnarray}
I_{\mathrm{off-spec}} &=& \frac{I_0}{h_i w_i} \frac {2 \pi \Delta\theta_{sc} {\cal A} r_e^2 {\vert t^{in}\vert}^2 {\vert t^{sc}\vert}^2 \left(\textbf{e}_{in}\cdot\textbf{e}_{sc}\right)^2} {k_0} \int_{-\infty}^{+\infty} dx e^{i q_{x_0} x} e^{-\frac{{\Delta q_x}^2 x^2}{2}} f\left(x, q_z \right),
\label{equation_pour_diffusion_offspeculaire}
\end{eqnarray}
which is the main result of this paper.
An efficient method for the numerical integration of Eq. (\ref{equation_pour_diffusion_offspeculaire}) is given in App. \ref{appaccordeon}.

\section{Examples}

\subsection{Experimental}
In this section we discuss several examples where the results from the previous sections are used to fit experimental data.
All the experiments reported here used a $27$ keV x-ray beam (wavelength $\lambda = 0.0459$ nm) at the CRG-IF beamline 
of the European Synchrotron Radiation Facility (ESRF). The scattering geometry is described in figure \ref{setup}. 
The monochromatic incident beam was first extracted from the polychromatic beam using a two-crystal Si(111) monochromator. 
Higher harmonics were eliminated using a W coated glass mirror, also used for focusing. In all experiments, 
the incident beam was $500\ \mu$m $\times 18\ \mu$m (W $\times$ H). The reflected intensity was defined using a 
$20$ mm $\times 200\ \mu$m (W $\times$ H) at $210$ mm from the sample and a $20$ mm $\times 200\ \mu$m (W $\times$ H) at $815$ mm 
from the sample and recorded using a NaI(Tl) scintillator.

Bilayers and double bilayers of $L{-}\alpha$ 1,2-distearoyl-sn-glycero-3-phosphocholine
(di-C$_{18}$-PC or DPSC) from Avanti Polar Lipids (Lancaster, Alabama, USA)
were prepared by first depositing a bilayer by two classical Langmuir-Blodgett (LB) depositions
(vertical sample).
We used super-polished ($< 1 \AA$ roughness) silicon substrates (SESO, Aix-en Provence, France) of surface 
$5 \times 5$ cm$^2$ and $1$ cm thick (to ensure planarity). 
The floating bilayer was then prepared by a LB deposition, followed by a Langmuir-Schaeffer (LS) deposition (horizontal sample)
\cite{charitat1999}.
The transfer rates are measured with 0.01 precision and 0.02 statistical dispersion.\\
Mixed OTS-lipid bilayers were prepared by first coating the substrate with an octadecyltrichlorosilane (OTS) 
as described in \cite{hughes2002}. Then, 2 monolayers of DSPC are deposited by Langmuir-Blodgett technique and 
a last one by Langmuir-Schaeffer technique. 

The samples are then inserted into a PTFE sample cell with $50 \mu$m thick windows
which is put in an aluminum box and thermalized using a water circulation bath. 
Sample are heated by steps with a feedback on the temperature measured inside the sample cell using a Pt100 resistance.
Specular reflectivity is obtained by rocking the sample for each angle of incidence ($q_x$ scans for approximately constant $q_z$ ) 
of incidence in order to subtract the background. Off-specular reflectivity is measured at a constant grazing angle of incidence 
of $0.4$ mrad bellow the critical angle to total external reflection at the Si-water interface ($0.7$ mrad). 
Background subtraction is described in detail in \cite{daillant2005}.

\subsection{Results and discussion}

\begin{figure}[h]
\begin{center}
\includegraphics[width=12cm]{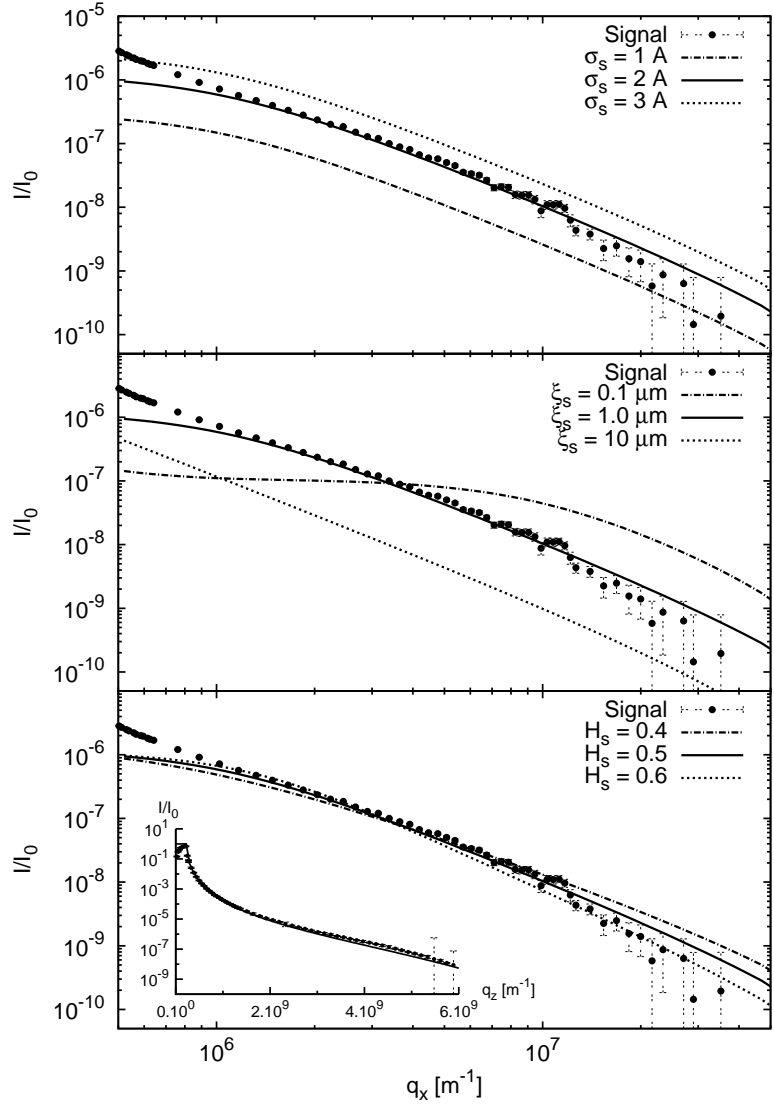}
\caption{\small{Top: effect of the average roughness $\sigma_s$ on the off-specular intensity. Middle: effect of the correlation length $\xi_s$ . Bottom: effect of the roughness exponent $H_s$. Inset: best fit of the specular reflectivity.}}
\label{diffus_silinu}
\end{center}
\end{figure}
We first determined the silicon substrate correlation function parameters, Eq. (\ref{Palazantsas}), by fitting the calculated specular 
and off-specular intensities to the experimental data for the substrate-water interface. 
Fig. \ref{diffus_silinu} shows the effect of these parameters. 
As expected, the off-specular scattering is more sensitive to the shape of the height-height correlation function than the 
specular reflectivity, which is in turn more sensitive to the electron density profile (thickness and density of the SiO$_2$ layer). 
The best set of parameters is $\sigma_s = 2 \pm 0.2 \AA$, $\xi_s = 1 \pm 0.1 \mu$m and $H_s = 0.5 \pm 0.05$, 
in good agreement with AFM characterization. Whereas a change in the roughness $\sigma_s$ results in an overall shift in
the scattered intensity in logarithmic scale, a change in the cutoff $\xi_s$ results in a shift in the crossover from a constant
intensity regime at low $q_x$ to a $1/q_x^{1+2H_s}$ regime at larger $q_x$ values.\\

\begin{figure}[h]
\begin{center}
\includegraphics[width=12cm]{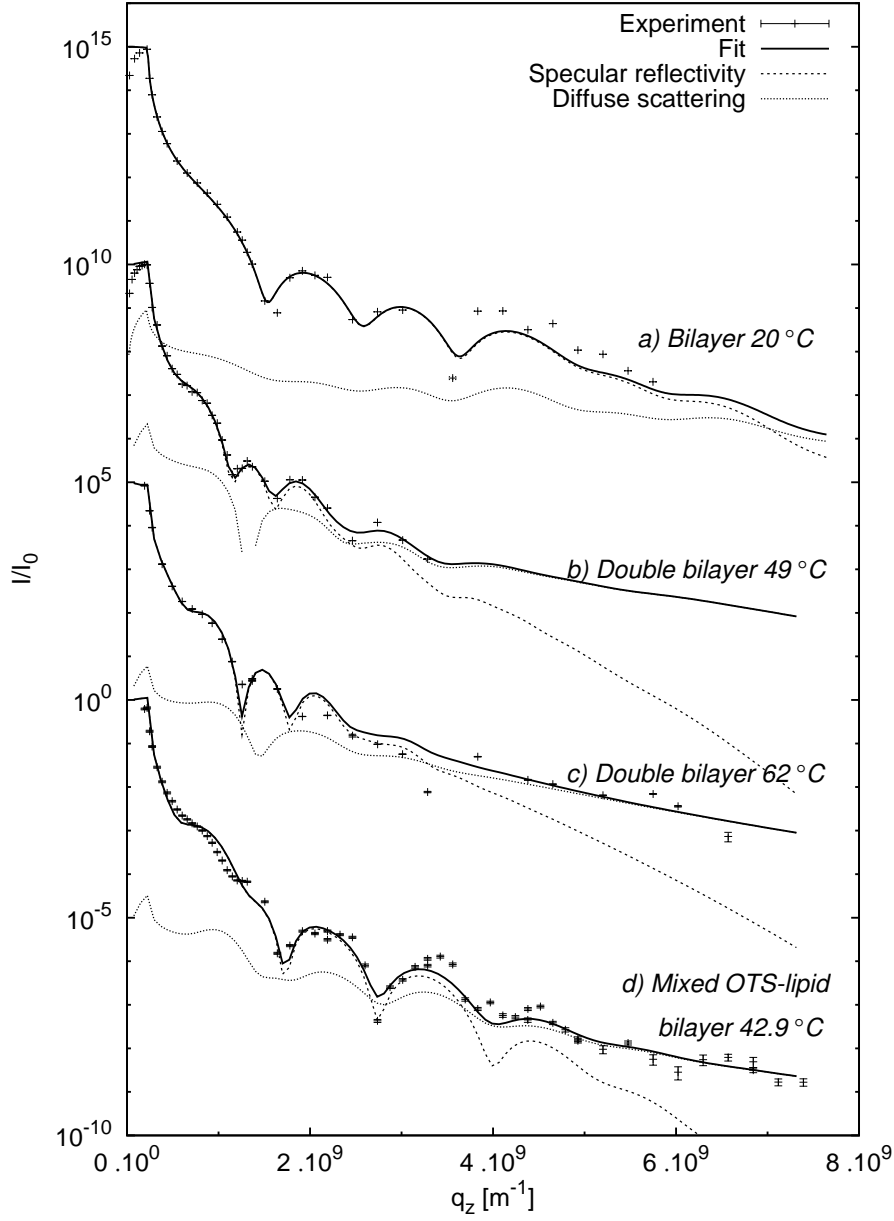}
\caption{Specular reflectivity and best fit (full line) for a supported bilayer at 20$^\circ$C (a), a double bilayer at 49$^\circ$C
in the gel phase (b), 62$^\circ$C in the fluid phase (c) and a mixed OTS-lipid bilayer at 42.9$^\circ$C in the gel phase (d).
The truely specular contribution is indicated using dashed lines and the contribution of the diffuse scattering in the specular direction 
is indicated using dotted line.}
\label{multispec}
\end{center}
\end{figure}
\begin{figure}
\begin{center}
\includegraphics[width=12cm]{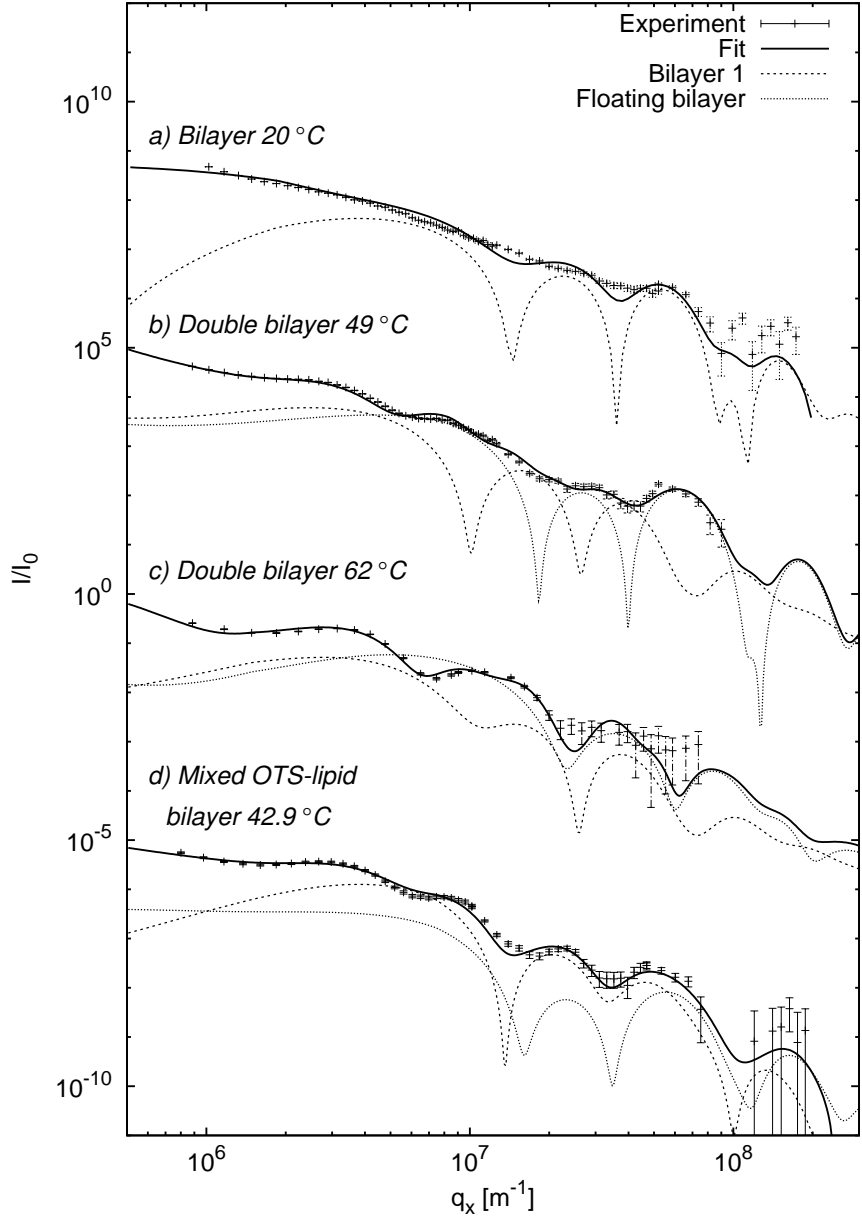}
\caption{Diffuse scattering and best fit (full line) for a supported bilayer at 20$^\circ$C (a), a double bilayer at 49$^\circ$C
in the gel phase (b), 62$^\circ$C in the fluid phase (c) and a mixed OTS-lipid bilayer at 42.9$^\circ$C in the gel phase (d).
The contribution of the first bilayer close to the substrate is indicated using dashed lines and the contribution of the floating bilayer
is indicated using dotted line.}
\label{multidiff}
\end{center}
\end{figure}
\begin{figure}
\begin{center}
\includegraphics[width=12cm]{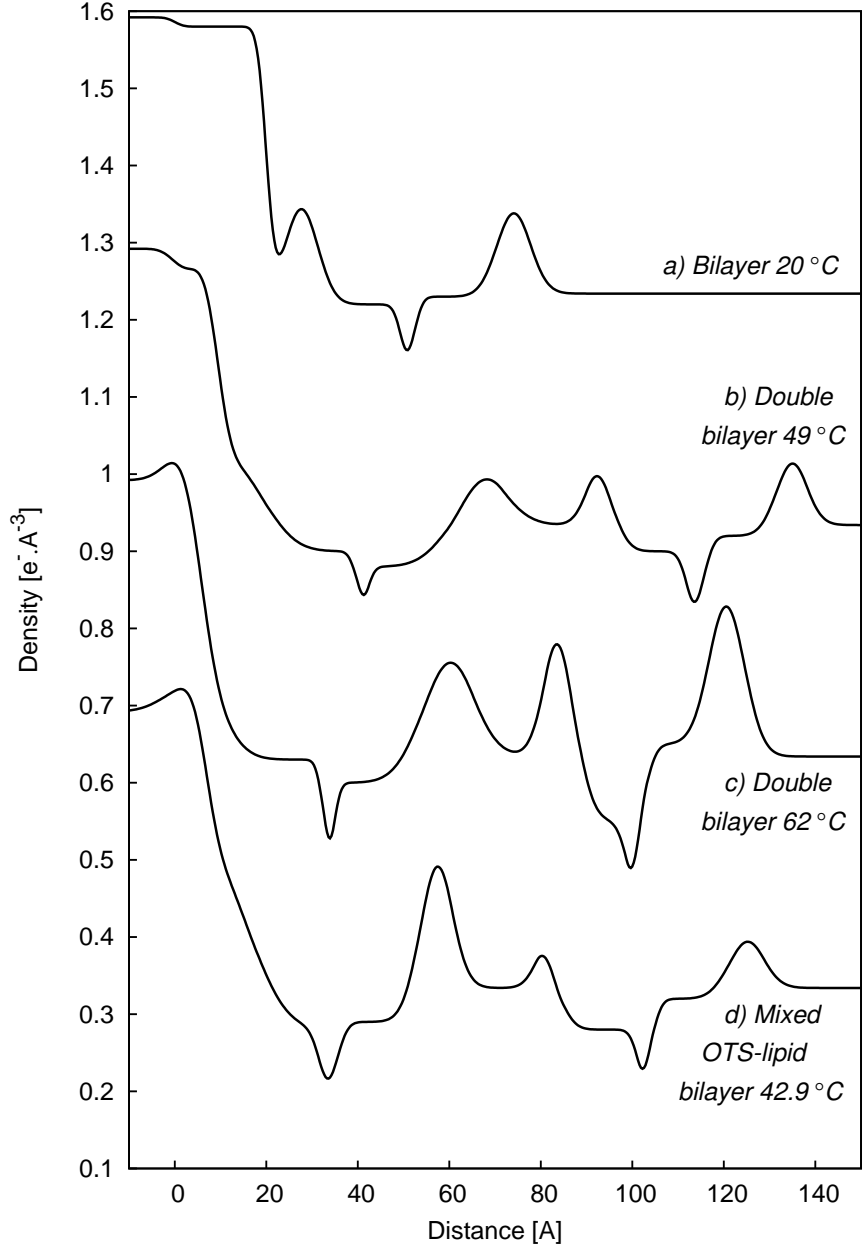}
\caption{\small{Electron-density profile of a di-C$_{18}$-PC obtained by combined specular and off-specular reflectivity using a 
Gaussian model comprising one Gaussian for the head groups and one Gaussian trough for the methyl groups \cite{Nagle89,Wiener89}.
(a) Bilayer, (b) double bilayer $49^\circ C$, (c) double bilayer $62^\circ C$, (c) mixed OTS-lipid bilayer $42.9^\circ C$.
}}
\label{multidens}
\end{center}
\end{figure}

The experimental results and best fits for off-specular and specular reflectivity are given in Figs. \ref{multispec} and
\ref{multidiff} for a single bilayer at $T = 20^{\circ}$C, a double bilayer at $T = 49^{\circ}$C and $T = 62^{\circ}$C
and a mixed OTS bilayer at $T = 42.9^{\circ}$C. 
Good fits could be obtained in all cases for both specular reflectivity and diffuse scattering.
In all cases, the off-specular scattering from the bare substrate provides a baseline for the contributions 
from the bilayers (Fig. \ref{multidiff}).
Electron densities which are represented on Fig. \ref{multidens} 
were modelled using the so-called 1-G gaussian model consisting in one gaussian for the head groups, a flat part describing
the chains and a Gaussian methyl trough \cite{Nagle89,Wiener89}. Parameters are given in  App. \ref{appsinglesupp}.

\begin{figure}
\begin{center}
\includegraphics[width=12cm]{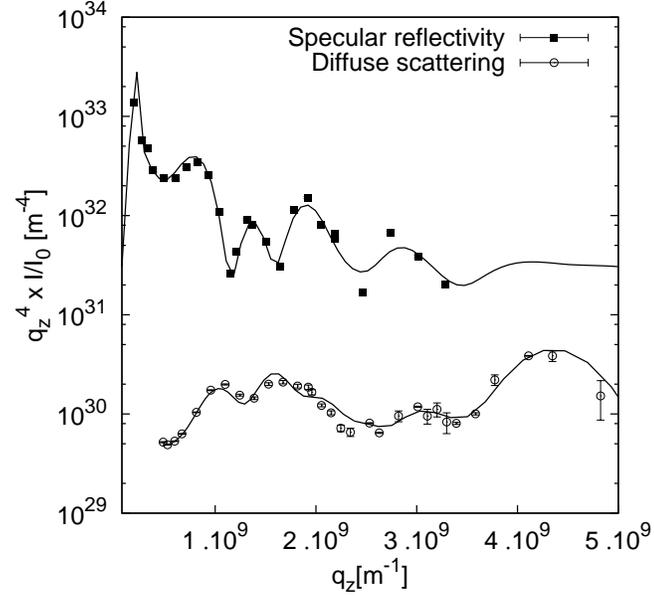}
\caption{Specular reflectivity and diffuse scattering from a double bilayer at 49$^\circ$C. 
$q_z^4 \times I/I_0$ as a function of $q_z$. The maxima in the specular intensity correspond to minima in the
diffuse scattering at low $q_z$.}
\label{compadiffspec}
\end{center}
\end{figure}

Whereas the specular curves (Fig. \ref{multispec}) are mainly sensitive to the average structure of the membrane, 
the off-specular signal (Fig. \ref{multidiff}) is both sensitive to the elastic properties of the bilayer and interaction
potentials, and to the structural parameters (because both $q_x$ and $q_z$ are varied in a detector scan, Fig. \ref{figsystemes}(b)).
In particular, whereas the shape of the specular signal results from interferences between
the $q_x \to 0$ components of the electron densities where all interfaces are fully correlated, 
the diffuse signal is directly determined by the strength of the correlations between interfaces 
at a given $q_x$, Eqs. (\ref{equation_pour_diffusion_offspeculaire}) and (\ref{fonctionf}), leading to a reduced contrast. 
This is demonstrated on Fig. \ref{compadiffspec}, where in addition to the reduced contrast, a contrast inversion 
between specular and diffuse reflectivities (maxima in the specular intensity correspond to minima in the diffuse scattering) 
can be seen at low $q_z$ as bilayer-bilayer correlations dominate over substrate-bilayer correlations. 
As a consequence, the substrate is not seen in low $q_z$ diffuse scattering, leading to the contrast inversion.

It should also be noted that the experimentally measured reflectivity signal consists of a truly specular signal and
a non-negligible diffuse contribution, both shown on Fig. \ref{multispec}, 
as the detector slits have a finite opening. The latter is relatively more important for large $q_z$ values as the specular
signal decays as $1/q_z^4 \exp(-q_z^2 \sigma^2)$, Eqs. (\ref{exprefl}), 
where $\sigma$ is a typical roughness value, whereas the off-specular signal decays as 
$1/q_x^{1+2H_s} \sim 1/q_z^{1/2+H_s}$,
Eqs. (\ref{equation_pour_diffusion_offspeculaire}) and (\ref{fonctionf}).\\
Due to the $\exp(-q_z^2 \sigma^2)$ decay in the specular signal, the contribution from the more strongly fluctuating floating bilayer 
to the specular intensity vanishes at large $q_z$. 
This is in contrast to the diffuse scattering where larger fluctuations make a larger contribution.
Accordingly the more strongly fluctuating floating bilayer makes the largest contribution 
at large $q_\parallel > 10^7 m^{-1}$ (see in particular the double bilayer in Fig. \ref{multidiff}), whereas the adsorbed bilayer 
and the substrate - adsobed bilayer cross-terms make a larger contribution at lower $q_\parallel$. 
Combining specular (Fig. \ref{multispec}) and off-specular (Fig. \ref{multidiff}) reflectivity is therefore 
essential in order to obtain a good sensitivity to the structural parameters of both bilayers (Fig. \ref{multidens}).

\begin{figure}[h]
	\begin{minipage}[b]{0.5\linewidth}
		\centering \includegraphics[width=8cm]{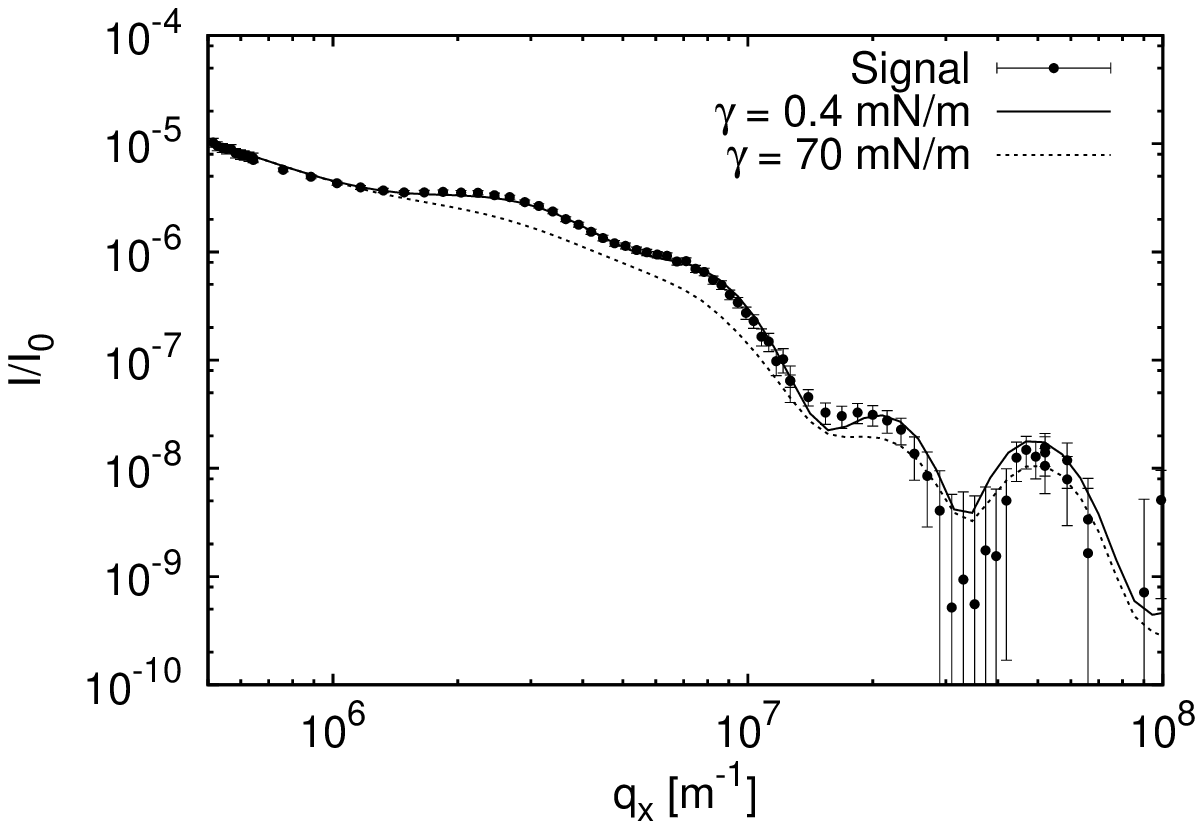}\\
		\centering \includegraphics[width=8cm]{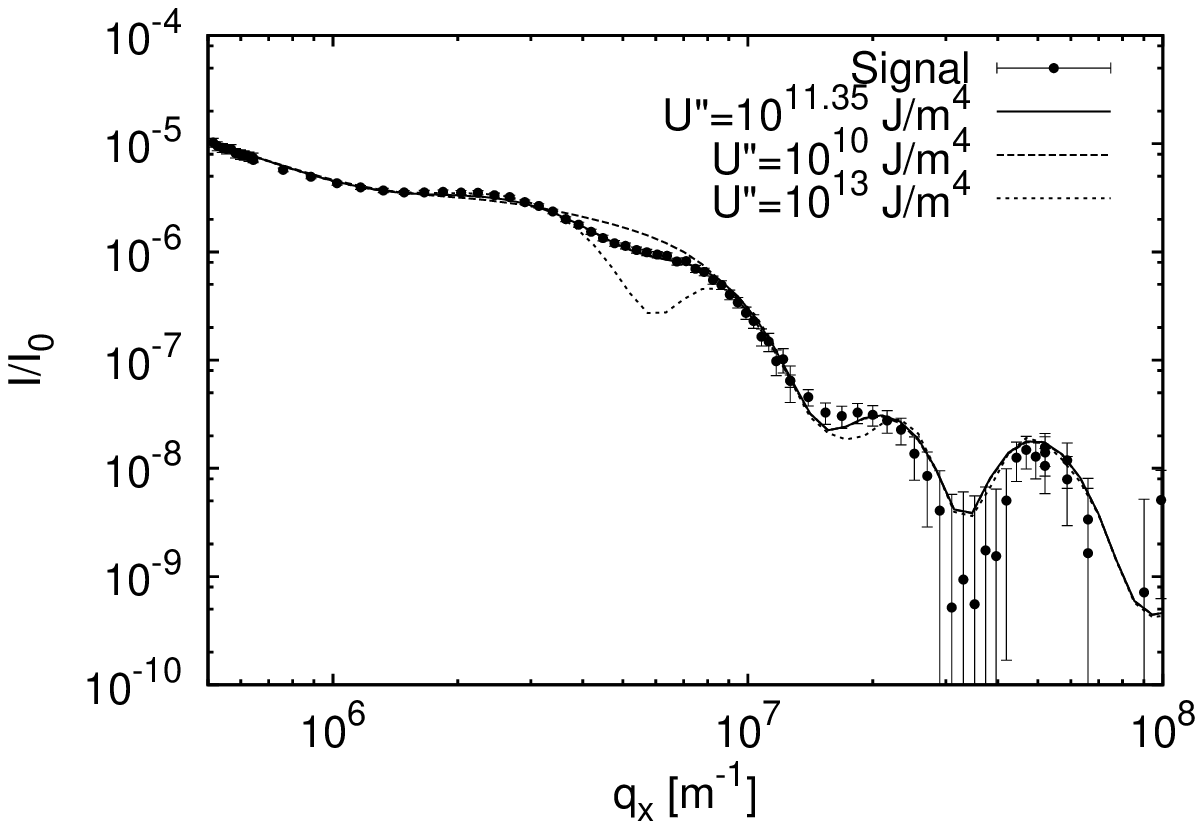}\\
	\end{minipage}\hfill
	\begin{minipage}[b]{0.5\linewidth}
		\centering \includegraphics[width=8cm]{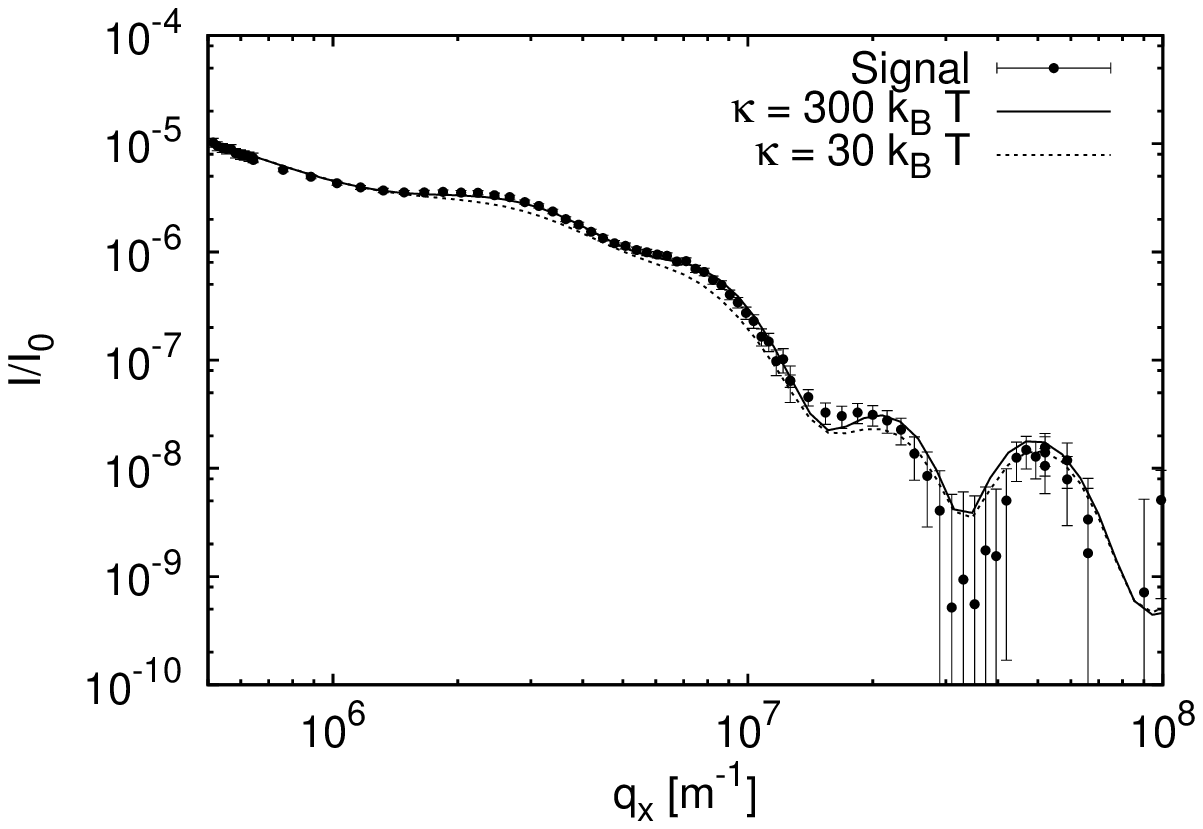}\\
		\centering \includegraphics[width=8cm]{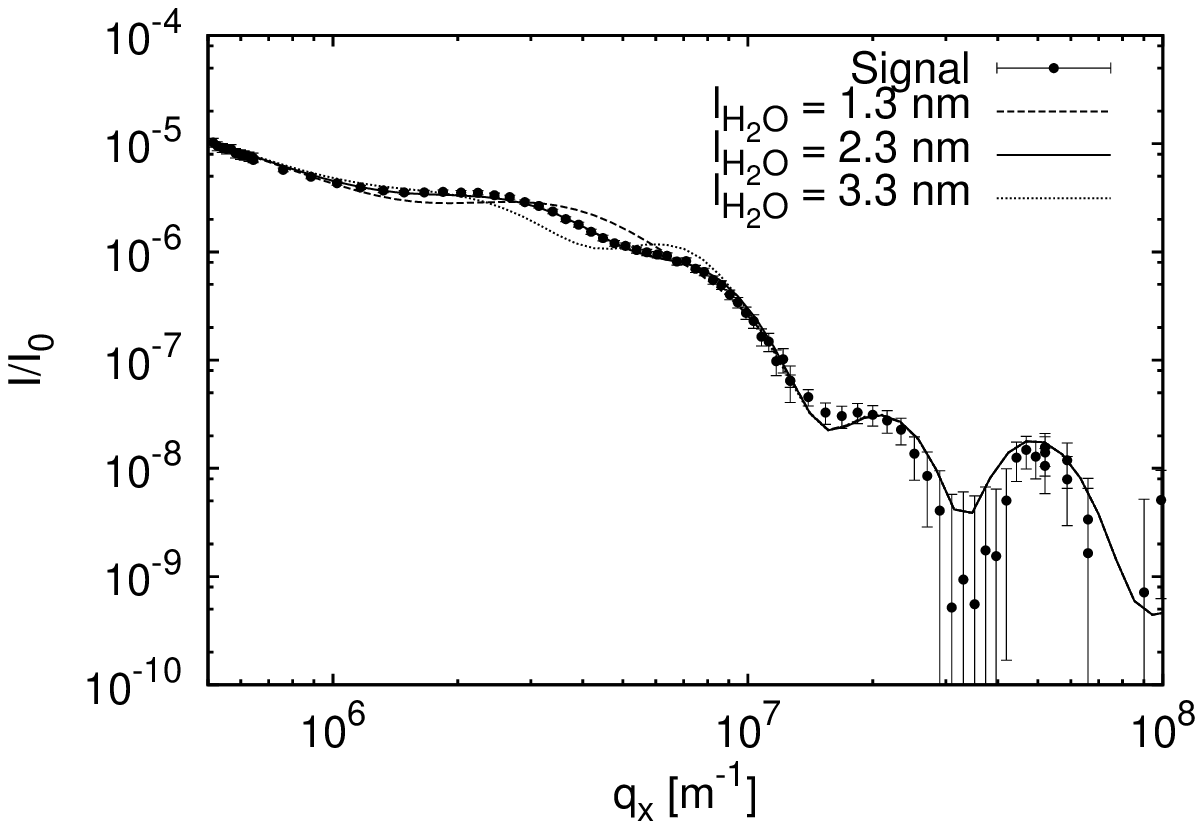}\\
	\end{minipage}
\caption{Effect of the elastic parameters of the floating bilayer, second derivative of the potential and water layer thickness on
the diffuse scattering from the mixed OTS-lipid bilayer at 42.9$^\circ$C. 
(a) Effect of the surface tension of the floating bilayer. (b) Effect of the bending rigidity. 
(c) Effect of the second derivative of the potential. (d) Effect of the thickness of the water layer.}
\label{graphes effet parametres ots}
\end{figure}

The sensitivity of the diffuse scattering to the tension, bending rigidity and second derivative of 
the potential is demonstrated on Fig.\ref{graphes effet parametres ots} for the mixed OTS-lipid sample.
Unless otherwize specified, on Fig.\ref{graphes effet parametres ots}, $\kappa = 300$ k$_B$ T 
which is a typical value for a lipid bilayer in the gel phase, 
$\gamma= 0.4$ mN/m, $U"=10^{11.35}$ J.m$^{-4}$ and $l_{H_2O} = 2.3$ nm which provide the best fit to the experimental curve.
A key region in the scattering curves is the fringe close to $q_\parallel \approx 5 \times 10^6$ m$^{-1}$ or
$q_z \approx 1.2\times 10^9$ m$^{-1} $, see also Fig. \ref{compadiffspec}, 
which results from the interference between x-rays scattered by the two bilayers. 
Its location is therefore extremely sensistive to the water layer thickness (Fig. \ref{graphes effet parametres ots}d) 
and its contrast is most sensitive to the second derivative of the potential 
(Fig. \ref{graphes effet parametres ots}c).
Too large a tension kills this correlation between the two bilayers.
However, the main effect of a smaller (resp. larger) tension is to increase (decrease) the scattering
at low $q_\parallel$ and decrease (increase) it at large $q_\parallel$ due to the larger (smaller) roughness
(Fig. \ref{graphes effet parametres ots}a).
Finally, the main effect of the bending rigidity is at large $q_\parallel$ (Fig. \ref{graphes effet parametres ots}b).
As all these parameters affect different regions of the curve in different ways, they are accurately determined 
in the fits without too much coupling.\\

Regarding the single bilayer, the best fit yielded $U" = 10^{11.2 \pm 0.5}$ J.m$^{-4}$ for the second derivative of the potential 
and a membrane tension $\gamma \approx 70 \pm 20$ mN.m$^{-1}$.
With this large $\gamma$ value, the fit is insensitive to the bending rigidity $\kappa$.
This large tension is much larger than lysis tension ($\approx 5-20$ mN.m$^{-1}$) \cite{evans(jphyschem1987)} and 
can be interpreted as an effective values describing lipid protusions rather than a usual tension.
Protusions are local independent motions of lipids, which should be the only possible motion for this strongly adsorbed membrane.
they tend to increase the local surface area and can therefore be described using an effective microscopic tension as suggested 
by Lipowsky and Grothehans \cite{Lipowsky1993}.
Lindahl et al. \cite{lindahl(biophys2000)} simulated the behavior of a floating bilayer found $\gamma_p = 50$ mN.m$^{-1}$ 
which is in the same order of magnitude as the surface tension we determine in our experiment. \\

Regarding double bilayers and mixed OTS bilayers, we obtain $\kappa \approx 200-300$ $k_B$ T in the gel phase 
which is in good agreement with the results found by P\'ecreaux et al on giant unilamellar vesicles \cite{pecreaux2004} 
and less than $50 k_B$T in the fluid phase.
Interestingly, we consistently obtain the effective tension of the first bilayer close to the substrate to be one order of magnitude
larger than that of the floating bilayer.
Whereas this can be attributed to protusion modes for double bilayers, a large effective tension 
is also not surprising for a mixed OTS-lipid bilayer where the OTS bilayer is polymerized and anchored on the substrate.
The tension of the floating membrane, $\gamma < 2$ mN/m, is larger than for vesicles but consistent 
with the fact that the membrane is still supported even if we consider it as a "floating" bilayer.
Regarding the second derivatives of the interaction potential, it is interesting to note that the smallest coupling between the 
first bilayer and the substrate is obtained for the mixed OTS-lipid bilayer, which is consistent with a polymerized OTS layer 
grafted at a relatively small number of points. On the other hand, the largest coupling is obtained for the adsorbed bilayer in the 
fluid phase. 
The second derivative of the interaction potential between the floating bilayer and the substrate is on the order 
of $(1-3)\times 10^{10} J.m^{-4}$ in the gel phase for either the double bilayer or the mixed OTS-lipid sample and 
decreases by more than one order of magnitude in the fluid phase. Finally, the largest interactions are between bilayers.\\


\section{Concluding remarks}
The formalism developed in this paper provides a rigourous frame for analyzing scattering data from 
supported bilayers. Coupling of specular and off-specular scattering is essential for determining the structure
of adsorbed and floating bilayers as well as their tension, rigidity and interaction potentials.
Several systems have been investigated, absorbed bilayers, double bilayers, mixed OTS-lipid bilayers.
Our data show that this new method gives a unique opportunity to investigate phenomena like protusion modes of adsorbed bilayers
and opens the way to the investigation of more complex systems like charged membranes or more realistic models 
including different kinds of lipids, cholesterol \cite{leslie09} or peptides.\\
More generally, our method based on a careful propagation of correlation functions provides an efficient 
scheme for tackling different systems like wetting films, polymer layers...where interaction potentials and/or
elastic parameters could be accurately determined.

Ackonowldgements: we wish to thank J.-S. Micha for assistance during the experiments, and G. Fragneto, F. Graner and
S. Lecuyer for discussions.
\clearpage
\section{Appendix}

\subsection{Fourier transform of Bessel functions}
\label{appfourier}

\begin{eqnarray}
\nonumber
\frac{1}{\left( 2 \pi \right)^2} \int d\textbf{q}_\parallel \frac{1}{q_\parallel^2 + q_{i \parallel }^2} e^{i \textbf{q}.{\bf r_\parallel}} &=& \frac{1}{2 \pi} \int_0^{+\infty} dq_\parallel  \frac{q_\parallel}{q_\parallel ^2 + q_{i \parallel }^2} J_0(q_\parallel r_\parallel)\\ 
&=& \frac{1}{2 \pi} K_0(q_{i \parallel} r_\parallel)\\
\nonumber
\frac{1}{\left( 2 \pi \right)^2} \int d\textbf{q}_\parallel \frac{1}{\left( q_\parallel ^2 + q_{i \parallel }^2 \right)^2} e^{i \textbf{q}_\parallel.{\bf r_\parallel}} & = & \frac{1}{2 \pi} \int_0^{+\infty} dq_\parallel  \frac{q_\parallel}{\left( q_\parallel^2 + q_{i \parallel}^2 \right)^2} J_0(q_\parallel r_\parallel)\\ 
&=& \frac{1}{2 \pi} \frac 12 q_{i \parallel} r_\parallel \frac{K_1(q_{i \parallel} r_\parallel)}{q_{i \parallel}^2}
\label{Bessel}
\end{eqnarray}

\subsection{Parameters for static correlation functions}

\subsubsection{Single bilayer}
\label{coefstatic1bic}
For a single supported bilayer $q_{1 \parallel}$ and $q_{2 \parallel}$ are the roots of the quadratic equation:
$$x^2 - \frac{\gamma_1}{\kappa_1} x + \frac{A_1}{\kappa_1} = 0.$$ 
Solving this equation numerically has to be done with care because of numerical instabilities and we follow the method of \cite{numrec} and set:
\begin{eqnarray}
s = - \frac 12 \left(-\frac{\gamma_1}{\kappa_1} - \sqrt{\frac{\gamma_1^2}{\kappa_1^2} - 4 \frac{A_1}{\kappa_1}}\right).
\end{eqnarray}
The roots of the equation are then given by:
\begin{eqnarray}
\nonumber
q_{1 \parallel} &=& s \\
\nonumber
q_{2 \parallel} &=& \frac{A_1}{\kappa_1 s},\\
\nonumber
\end{eqnarray}
and the coefficients $\lambda_i, \eta_i$ are given by:
\begin{eqnarray}
\lambda_1 &=& \frac{2}{\left(q_{1 \parallel}^2-q_{2 \parallel}^2\right)^3} \mathrm{\hspace{0.5cm} and \hspace{0.5cm}} \lambda_2 = - \lambda_1 \\
\eta_1 &=& -\frac{1}{q_{1 \parallel}^2-q_{2 \parallel}^2} \mathrm{\hspace{0.5cm} and \hspace{0.5cm}} \eta_2 = - \eta_1.
\end{eqnarray}

\subsubsection{Double bilayer}
\label{coefstatic2bic}
For a double  supported bilayer we need to solve a quartic equation leading to $q_{i \parallel}$ (i=1...4). Again, we use the method 
described in \cite{numrec}. The static coefficients $ \nu_i$ and $\eta_{i,j}$ are given by:
\begin{eqnarray}
\nu_{j} &=&  \frac{\left( \delta_{1,1}^2 - q_{j \parallel}^2 \right) \left( \delta_{1,2}^2 - q_{j \parallel}^2 \right)\left( \delta_{2,1}^2 - q_{j \parallel}^2 \right) \left( \delta_{2,2}^2 - q_{j \parallel}^2 \right)}{\prod_{k \neq j} (q_{k \parallel}^2 - q_{j \parallel}^2)^2}\\
\eta_{i,j} &=&  \frac{\left( \delta_{i,1}^2 - q_{j \parallel}^2 \right) \left( \delta_{i,2}^2 - q_{j \parallel}^2 \right)}{\prod_{k \neq j} (q_{k \parallel}^2 - q_{j \parallel}^2)},
\end{eqnarray}
where $\delta_{1,1}$ and $\delta_{1,2}$ (resp. $\delta_{2,1}$ and $\delta_{2,2}$) are the solutions of 
\begin{eqnarray}
\nonumber
X^2 - \frac{\gamma_2}{\kappa_2} X + \frac{A_1 A_2 +B (A_1 + A_2)}{A_1 \kappa_2}&=&0
\end{eqnarray}
and
\begin{eqnarray}
\nonumber
X^2 - \frac{\gamma_1}{\kappa_1} X + \frac{A_1 A_2 +B (A_1 + A_2)}{A_2 \kappa_1}&=&0.
\end{eqnarray}

The expression for $\tau_i$ and $\lambda_{i,j}$ are more complex and will not be given here.

\subsection{Diagonalization of the free energy for a double supported bilayer}
\label{appaigen}
We have:
\begin{equation}
\nonumber
{\cal F}_{q_\parallel} = \left(\begin{array}{cc} \tilde{u}_1(-q_\parallel) & \tilde{u}_2(-q_\parallel)\end{array}\right) \left(\begin{array}{cc} \frac 12 \left( \tilde{a}_1(q_\parallel) + B \right) & -B/2 \\ -B/2 &  \frac 12 \left( \tilde{a}_2(q_\parallel) + B \right) \end{array}\right) \left(\begin{array}{c} \tilde{u}_1(q_\parallel) \\ \tilde{u}_2(q_\parallel)\end{array}\right).
\end{equation}
The diagonalization of the matrix ${\bar{\cal F}}$ gives the eigenvalues:
\begin{eqnarray}
\nonumber
\lambda_+ (q_\parallel) &=& \frac 14 \left( \tilde{a}_1(q_\parallel) + \tilde{a}_2(q_\parallel) + 2B + \sqrt{\Delta(q_\parallel)} \right)\\
\lambda_- (q_\parallel) &=& \frac 14 \left( \tilde{a}_1(q_\parallel) + \tilde{a}_2(q_\parallel) + 2B - \sqrt{\Delta(q_\parallel)} \right),
\end{eqnarray}
and the eigenvectors:
\begin{eqnarray}
\nonumber
{U}_+ &=& \frac{B}{\Phi(q_\parallel)} \left( - \frac{\tilde{a}_1(q_\parallel) - \tilde{a}_2(q_\parallel) + \sqrt{\Delta(q_\parallel)}}{2B},1 \right) \\
{U}_- &=& \frac{B}{\Psi(q_\parallel)} \left( - \frac{\tilde{a}_1(q_\parallel) - \tilde{a}_2(q_\parallel) - \sqrt{\Delta(q_\parallel)}}{2B},1 \right),
\end{eqnarray}
with:
\begin{eqnarray}
\nonumber
\Delta(q_\parallel) &=& \left(\tilde{a}_1(q_\parallel) - \tilde{a}_2(q_\parallel)\right)^2 + 4B^2\\
\nonumber
\Phi(q_\parallel) &=& \frac 12 \sqrt{4B^2 + \left(\tilde{a}_1(q_\parallel)-\tilde{a}_2(q_\parallel) + \sqrt{\Delta(q_\parallel)}\right)^2}\\
\Psi(q_\parallel) &=& \frac 12 \sqrt{4B^2 + \left(\tilde{a}_1(q_\parallel)-\tilde{a}_2(q_\parallel) - \sqrt{\Delta(q_\parallel)}\right)^2}.
\end{eqnarray}
In the new basis,
\begin{eqnarray}
{\cal F}_{q_\parallel} &=& \lambda_+ (q_\parallel) g^2(q_\parallel) + \lambda_- (q_\parallel) h^2(q_\parallel),
\end{eqnarray}
with:
\begin{eqnarray}
\nonumber
g(q_\parallel) &=& \frac{- \left( \tilde{a}_1(q_\parallel) - \tilde{a}_2(q_\parallel) + \sqrt{\Delta(q_\parallel)} \right) \tilde{u}_1(q_\parallel) + 2B \tilde{u}_2(q_\parallel)}{2 \Phi(q_\parallel)}\\
h(q_\parallel) &=& \frac{- \left( \tilde{a}_1(q_\parallel) - \tilde{a}_2(q_\parallel) - \sqrt{\Delta(q_\parallel)} \right) \tilde{u}_1(q_\parallel) + 2B \tilde{u}_2(q_\parallel)}{2 \Psi(q_\parallel)}.
\label{modespropres}
\end{eqnarray}
We can now apply the theorem of equipartition of energy,
\begin{eqnarray}
\nonumber
\left< |g(q_\parallel)|^2 \right> = \frac{k_B T}{2 \lambda_+}\\
\left< |h(q_\parallel)|^2 \right> = \frac{k_B T}{2 \lambda_-}.
\label{equipartition}
\end{eqnarray}
$\tilde{u}_1(q_\parallel)$ and $\tilde{u}_2(q_\parallel)$ can now be calculated by inverting Eq. (\ref{modespropres}).
\begin{eqnarray}
\nonumber
\tilde{u}_1(q_\parallel) &=& k_B T \frac{\Phi(q_\parallel) h(q_\parallel) - \Psi(q_\parallel) g(q_\parallel)}{2 \sqrt{\Delta(q_\parallel)}}\\
\tilde{u}_2(q_\parallel) &=& k_B T \frac{\left( \tilde{a}_1(q_\parallel) - \tilde{a}_2(q_\parallel) + \sqrt{\Delta(q_\parallel)} \right) \Psi(q_\parallel) h(q_\parallel) - \left( \tilde{a}_1(q_\parallel) - \tilde{a}_2(q_\parallel) - \sqrt{\Delta(q_\parallel)}\right) \Phi(q_\parallel) g(q_\parallel)} {4 B \sqrt{\Delta(q_\parallel)}}.
\end{eqnarray}

\subsection{Coefficients of the thermal correlation functions}

\subsubsection{Single bilayer}
\label{thermocoef1bic}

In the single bilayer case, coefficients of the thermal correlation functions  are simply given by:

\begin{eqnarray}
\alpha_{1} = \frac{1}{q_{2 \parallel}^2 - q_{1 \parallel}^2}, \alpha_{2} = - \frac{1}{q_{2 \parallel}^2 - q_{1 \parallel}^2}.
\label{alphai}
\end{eqnarray}

\subsubsection{Double bilayer}
\label{thermocoef2bic}

In the double bilayer case, coefficients of the thermal correlation functions  are given by:

\begin{eqnarray}
\alpha_{1,j} &=&  \frac{\left( \beta_{2,1}^2 - q_{j \parallel}^2 \right) \left( \beta_{2,2}^2 - q_{j \parallel}^2 \right)}{\prod_{k \neq j} (q_{k \parallel}^2 - q_{j \parallel}^2)}\\
\alpha_{2,j} &=&  \frac{\left( \beta_{1,1}^2 - q_{j \parallel}^2 \right) \left( \beta_{1,2}^2 - q_{j \parallel}^2 \right)}{\prod_{k \neq j} (q_{k \parallel}^2 - q_{j \parallel}^2)}\\
\iota_{j} &=&  \frac{1}{\prod_{k \neq j} (q_{k \parallel}^2 - q_{j \parallel}^2)}
\end{eqnarray}

where $\beta_{i,j}^2$ are the roots of equations:

\begin{eqnarray}
x^2 - \frac{\gamma_i}{\kappa_i} x + \frac{A_i+B}{\kappa_i} = 0.
\end{eqnarray}

\subsection{Integration of Eq. (\ref{equation_pour_diffusion_offspeculaire}) for off-specular scattering}
\label{appaccordeon}

When numerically integrating Eq. (\ref{equation_pour_diffusion_offspeculaire})
$\cos(q_{x_0}x)$ oscillates very fast whereas $f\left(x, q_z \right)$ decays logarithmically. 
Setting $u = q_{x_0} x$ and next $u = v+2k\pi$ with $v \in \left[ 0 ;\pi/2 \right]$ we obtain:
\begin{eqnarray}
\nonumber
I_{\mathrm{off-spec}} &=& \frac{I_0}{h_i w_i} \frac {4 \pi \Delta\theta {\cal A} r_e^2 {\vert t^{in}\vert}^2 {\vert t^{sc}\vert}^2 \left(\textbf{e}_{in}\cdot\textbf{e}_{sc}\right)^2} {k_0}  \frac{1}{q_{x_0}} \displaystyle { \sum_{k=0}^{N}} \left[ \int_{0}^{\pi/2} dv \cos\left(v\right) e^{-\frac{{\Delta q_x}^2 \left(v +2 k \pi\right)^2}{2 q_{x_0}^2}} f\left(\frac{v +2 k \pi}{q_{x_0}}, q_z \right) \right. \\
\nonumber
&+& \left. \int_{\pi/2}^{\pi} dv \cos\left(v + \frac{\pi}{2}\right) e^{-\frac{k_0^2 {\sin\theta}^2 {\Delta\theta}^2 \left(v +2 k \pi + \pi/2\right)^2}{2 q_{x_0}^2}} f\left(\frac{v +2 k \pi + \pi/2}{q_{x_0}}, q_z \right) \right. \\
\nonumber
&+& \left. \int_{\pi}^{3 \pi/2} dv \cos\left(v + \pi\right) e^{-\frac{{\Delta q_x}^2 \left(v +2 k \pi + \pi\right)^2}{2 q_{x_0}^2}} f\left(\frac{v +2 k \pi + \pi}{q_{x_0}}, q_z \right) \right. \\
\nonumber
&+& \left. \int_{3 \pi/2}^{2 \pi} dv \cos\left(v + \frac{3 \pi}{2}\right) e^{-\frac{{\Delta q_x}^2 \left(v +2 k \pi + 3 \pi/2\right)^2}{2 q_{x_0}^2}} f\left(\frac{v +2 k \pi + 3 \pi / 2}{q_{x_0}}, q_z \right) \right].\\
\label{intensite_diffusee}
\end{eqnarray}
$N = 100$ ensures fast convergence.

\subsection{Structural parameters}
\label{appsinglesupp}

\begin{table}[h]                                                                                                                                                                 
\begin{center}                                                                                                                                                                   
\begin{tabular}{|c||c c c c c|}
\hline
                              & bare            & single                & double                & double               & OTS mixed               \\
                              & substrate       & bilayer ($20^\circ$C) & bilayer ($49^\circ$C) & bilayer ($62^\circ$C)& bilayer ($42.9^\circ$C) \\
\hline
\hline
$\rho_{Si} [e^-/\AA^3]$       & $0.692$         & $0.692$               & $0.692$               &  $0.692$             & $0.692$                 \\
$\sigma_{Si} [\AA]$           & $2 \pm 0.2$     & $1.9 \pm 0.2$         & $4.6 \pm 0.2$         &  $3.9 \pm 0.2$       & $4 \pm 0.2$             \\
$\rho_{Si0_2} [e^-/\AA^3]$    & $0.52 \pm 0.05$ & $0.52 \pm 0.05$       & $0.65 \pm 0.05$       &  $0.73 \pm 0.05$     & $0.66 \pm 0.02$         \\
$D_{Si0_2} [\AA]$             & $3.8 \pm 0.5$   & $3.8 \pm 0.5$         & $9.3 \pm 0.5$         &  $5.1 \pm 0.5$      & $6.5 \pm 0.5$           \\
$\sigma_{Si0_2} [\AA]$        & $2 \pm 0.2$     & $1.9 \pm 0.2$         & $3.6 \pm 0.2$         &  $4.8 \pm 0.2$      & $4 \pm 0.2$             \\
$\xi_s [\mu m]$               & $1 \pm 0.5$     & $1 \pm 0.5$           & $6.8 \pm 0.5$         &  $2.8 \pm .5$        & $1 \pm 0.1$             \\
$H_s$                         & $0.5 \pm 0.05$  & $0.5 \pm 0.05$        & $0.43 \pm 0.05$       &  $0.45 \pm 0.05$     & $0.38 \pm 0.05$         \\
$D_{1,H_2O} [\AA]$            &                 & $7 \pm 1$             & $5.5 \pm 1$           &  $1.5 \pm 1$         & $0.5 \pm 1$             \\
$\rho_{1,head} [e^-/\AA^3]$   &                 & $0.45 \pm 0.02$       & $0.42 \pm 0.02$       &  $0.40 \pm 0.02$     & $0.5 \pm 0.02$          \\
$\rho_{1,tail} [e^-/\AA^3]$   &                 & $0.33 \pm 0.02$       & $0.30 \pm 0.02$       &  $0.32 \pm 0.02$     & $0.31 \pm 0.02$         \\
$\rho_{CH_3,B_1} [e^-/\AA^3]$ &                 & $0.26 \pm 0.02$       & $0.28 \pm 0.02$       &  $0.25 \pm 0.02$     & $0.26 \pm 0.02$         \\
$\rho_{2,tail} [e^-/\AA^3]$   &                 & $0.33 \pm 0.02$       & $0.29  \pm 0.02$      &  $0.30 \pm 0.02$     & $0.29 \pm 0.02$         \\
$\rho_{2,head} [e^-/\AA^3]$   &                 & $0.44 \pm 0.02$       & $0.41 \pm 0.02$       &  $0.46 \pm 0.02$     & $0.51 \pm 0.02$         \\
$d_{1,head} [\AA]$            &                 & $3.8 \pm 0.5$           & $6.1 \pm 0.5$         &  $5.1 \pm 0.5$       & $7 \pm 0.5$             \\
$D_{1,tail} [\AA]$            &                 & $22.5 \pm 0.5$        & $26.0 \pm 0.5$        &  $26.9 \pm 0.5$      & $22.7 \pm 0.5$          \\
$d_{CH_3,B_1} [\AA]$          &                 & $2.0 \pm 0.5$         & $1.5  \pm 0.5$        &  $1.75 \pm 0.5$      & $2.1 \pm 0.5$           \\
$D_{2,tail} [\AA]$            &                 & $23.0 \pm 0.5$        & $24.5 \pm 0.5$        &  $25.7 \pm 0.5$      & $21 \pm 0.5$            \\
$d_{2,head} [\AA]$            &                 & $4.3 \pm 0.5$           & $6.1 \pm 0.5$         &  $5.7 \pm 0.5$       & $3.5 \pm 0.5$           \\
$D_{2,H_2O} [\AA]$            &                 &                       & $28.0 \pm 1$          &  $25.6 \pm 1$        & $25.1 \pm 1$            \\
$\rho_{3,head} [e^-/\AA^3]$   &                 &                       & $0.41 \pm 0.02$       &  $0.40 \pm 0.02$     & $0.39 \pm 0.02$         \\
$\rho_{3,tail} [e^-/\AA^3]$   &                 &                       & $0.29 \pm 0.02$       &  $0.25 \pm 0.02$     & $0.28\pm 0.02$          \\
$\rho_{CH_3,B_2} [e^-/\AA^3]$ &                 &                       & $0.26 \pm 0.02$       &  $0.24 \pm 0.02$     & $0.26 \pm 0.02$         \\
$\rho_{4,tail} [e^-/\AA^3]$   &                 &                       & $0.31 \pm 0.02$       &  $0.33 \pm 0.02$     & $0.32 \pm 0.02$         \\
$\rho_{4,head} [e^-/\AA^3]$   &                 &                       & $0.42 \pm 0.02$       &  $0.41 \pm 0.02$     & $0.40 \pm 0.02$         \\
$d_{3,head} [\AA]$            &                 &                       & $3.1 \pm 0.5$         &  $3.5 \pm 0.5$       & $2.9 \pm 0.5$           \\
$D_{3,tail} [\AA]$            &                 &                       & $20.7 \pm 0.5$        &  $16.3 \pm 0.5$      & $21.0 \pm 0.5$          \\
$d_{CH_3,B_2} [\AA]$          &                 &                       & $2.1 \pm 0.5$         &  $2.2 \pm 0.5$       & $1.8 \pm 0.5$           \\
$D_{4,tail} [\AA]$            &                 &                       & $20.6 \pm 0.5$        &  $19.6 \pm 0.5$      & $21.6 \pm 0.5$          \\
$d_{4,head} [\AA]$            &                 &                       & $3.4 \pm 0.5$         &  $3.9 \pm 0.5$       & $3.9 \pm 0.5$           \\
$\gamma_1 [mN/m]$             &                 & $70 \pm 10$           & $15.0 \pm 1$          &  $5.0 \pm 0.2$       & $15.8 \pm 2$            \\
$\kappa_1 [k_BT]$             &                 & $350 \pm 100$         & $370 \pm 50$          &  $110 \pm 25$        & $350 \pm 100$           \\
$Log(U''_{1,s} [J.m^{-4}])$   &                 & $11.2 \pm 1 $         & $10.3 \pm 1 $         &  $10.9 \pm 1$        & $10.3 \pm 1$            \\
$\gamma_2 [mN/m]$             &                 &                       & $1.0 \pm 1$           &  $0.09 \pm 0.1$      & $0.1 \pm 0.05$          \\
$\kappa_2 [k_BT]$             &                 &                       & $150 \pm 50$          &  $45  \pm 25$        & $200 \pm 50$            \\
$Log(U''_{2,s} [J.m^{-4}])$   &                 &                       & $8.9 \pm 1 $          &  $8.3 \pm 1$         & $10.0 \pm 1$            \\
$Log(U''_{1,2} [J.m^{-4}])$   &                 &                       & $12.2 \pm 0.3 $       &  $13.4 \pm 1$        & $12.2 \pm 1$            \\ 
\hline
\end{tabular}
\end{center}
\caption{$\rho$: electron density; $D$: box thickness; $d$: Gaussian width; $\sigma$: roughness; $\xi_s$:cutoff, $H_s$: roughness exponent,
$\gamma$: tension, $\kappa$: bending rigidity, $U''$: second derivative of the interaction potential. Subscript s represents substrate, 1 membrane 1 
(next to the substrate), and 2 membrane 2.}
\end{table}

\bibliographystyle{unsrt}
\bibliography{bibliomembranes}

\newcommand{\SortNoop}[1]{}
\begin{thebibliography}{10}

\bibitem{pecreaux2004}
J.~P\'ecr\'eaux, D\"obereiner~H. G., J.~Prost, J.~F. Joanny, and P.~Bassereau.
\newblock Refined contour analysis of giant unilamellar vesicles.
\newblock {\em European Physical Journal E}, 13:277--290, 2004.

\bibitem{Brzustowicz:ko5011}
M.~R. Brzustowicz and A.~T. Brunger.
\newblock {X-ray scattering from unilamellar lipid vesicles}.
\newblock {\em Journal of Applied Crystallography}, 38(1):126--131, Feb 2005.

\bibitem{petrache(pre1998)}
H.~I. Petrache, N.~Gouliaev, S.~Tristram-Nagle, S.~Zhang, R.~M. Suter, and
  J.~F. Nagle.
\newblock Interbilayer interactions from high-resolution x-ray scattering.
\newblock {\em Physical Review E}, 57:7014--7024, 1998.

\bibitem{petrache(biophysj2004)}
H.~I Petrache, S.~Tristram-Nagle, K.~Gawrisch, D.~Harries, V.~A. Parsegian, and
  J.~F. Nagle.
\newblock Structure and fluctuations of charged phosphatidylserine bilayers in
  the absence of salt.
\newblock {\em Biophysical Journal}, 86:1574--1586, 2004.

\bibitem{NagleBBA2000}
J.F. Nagle and S.~Tristram-Nagle.
\newblock Structure of lipid bilayers.
\newblock {\em BBA Biomembranes}, 1469:159--195, 2000.

\bibitem{caille72}
A.~Caill\'e.
\newblock Remarques sur la diffusion des rayons x dans les smectiques a.
\newblock {\em Comptes Rendus de l'AcadŽmie des Sciences Paris}, pages
  891--893, 1972.

\bibitem{Nall89}
J.~F. Nallet, D.~Roux, and J.~Prost.
\newblock Dynamic light scattering study of dilute lamellar phases.
\newblock {\em Physical Review Letters}, 62:276--279, 1989.

\bibitem{rand(bba1989)}
R.~P. Rand and V.~A. Parsegian.
\newblock Hydration forces between phospholipid bilayers.
\newblock {\em Biochimica et Biophysica Acta}, 988:351--376, 1989.

\bibitem{Salditt2000}
M.~Vogel, C.~Munster, W.~Fenzl, and T.~Salditt.
\newblock Thermal unbinding of highly oriented phospholipid membranes.
\newblock {\em Physical Review Letters}, 84:390--393, 2000.

\bibitem{Jeu96}
W.~H.~De Jeu, J.~D. Shindler, and E.~A.~L. Mol.
\newblock The resolution function in diffuse x-ray reflectivity.
\newblock {\em J. Applied Cryst.}, 29:511--515, 1996.

\bibitem{brian(pnas1984)}
A.~A. Brian and H.~M. McConnell.
\newblock {Allogeneic stimulation of cytotoxic T cells by supported planar
  membranes}.
\newblock {\em Proceedings of the National Academy of Sciences of the United
  States of America}, 81(19):6159--6163, 1984.

\bibitem{charitat1999}
T.~Charitat, E.~Bellet-Amalric, G.~Fragneto, and F.~Graner.
\newblock Adsorbed and free lipid bilayers at the solid-liquid interface.
\newblock {\em European Physical Journal B}, 8:583--593, 1999.

\bibitem{Giocondi2004861}
M-C. Giocondi, P.~E. Milhiet, P.~Dosset, and C.~Le~Grimellec.
\newblock Use of cyclodextrin for afm monitoring of model raft formation.
\newblock {\em Biophysical Journal}, 86(2):861 -- 869, 2004.

\bibitem{nickel08}
Christian Reich, Margaret~R. Horton, Baerbel Krause, Alice~P. Gast, Joachim~O.
  Raedler, and Bert Nickel.
\newblock {Asymmetric structural features in single supported lipid bilayers
  containing cholesterol and G(M1) resolved with synchrotron x-ray
  reflectivity}.
\newblock {\em {Biophysical Journal}}, {95}({2}):{657--668}, {2008}.

\bibitem{miller(PRL2008)}
C.~E. Miller, J.~Majewski, E.~B. Watkins, D.~J. Mulder, T.~Gog, and T.~L. Kuhl.
\newblock Probing the local order of single phospholipid membranes using
  grazing incidence x-ray diffraction.
\newblock {\em Physical Review Letters}, 100:058103, 2008.

\bibitem{Wagner2000}
M.~L. Wagner and L.~K. Tamm.
\newblock Tethered polymer-supported planar lipid bilayers for reconstitution
  of integral membranes: silane-polyethyleneglycol-lipid as a cushion and
  covalent linker.
\newblock {\em Biophysical Journal}, 79:1400--1414, 2000.

\bibitem{Sinner2001}
E.~Sinner and W.~Knoll.
\newblock Functional tethered membranes.
\newblock {\em Curr. Op. Chem. Biol.}, 5:705--711, 2001.

\bibitem{Beerlink2008}
A.~Beerlink, P.-J. Wilbrandt, E.~Ziegler, D.~Carbone, T.~H. Metzger, and
  T.~Salditt.
\newblock X-ray structure analysis of free-standing lipid membranes facilitated
  by micromachined apertures.
\newblock {\em Langmuir}, 24(9):4952--4958, 2008.

\bibitem{hughes2002}
A.~V. Hughes, A.~Goldar, M.~C. Gestenberg, S.~J. Roser, and J.~Bradshaw.
\newblock A hybrid sam phospholipid approach to fabricating a free supported
  lipid bilayer.
\newblock {\em Physical Chemistry Chemical Physics}, 4:2371--2378, 2002.

\bibitem{daillant2005}
J.~Daillant, E.~Bellet-Amalric, A.~Braslau, T.~Charitat, G.~Fragneto,
  F.~Graner, S.~Mora, F.~Rieutord, and B.~Stidder.
\newblock Structure and fluctuations of a single floating lipid bilayer.
\newblock {\em The Proceding of the National Academy of Sciences USA},
  102:11639--11644, 2005.

\bibitem{andelman1999}
P.~S. Swain and D.~Andelman.
\newblock The influence of substrate structure on membrane adhesion.
\newblock {\em Langmuir}, 15:8902--8914, 1999.

\bibitem{andelman2001}
P.~S. Swain and D.~Andelman.
\newblock Supported membranes on chemically structured and rough surfaces.
\newblock {\em Physical Review E}, 63:51911, 2001.

\bibitem{merath07}
R.~J. Merath and U.~Seifert.
\newblock Fluctuation spectra of free and supported membrane pairs.
\newblock {\em Eur. Phys. J. E}, 23(1):103--116, may 2007.

\bibitem{pershan(1991)}
I.~M. Tidswell, T.~A. Rabedeau, P.~S. Pershan, and S.~D. Kosowsky.
\newblock Complete wetting of a rough surface: an x-ray study.
\newblock {\em Physical Review Letters}, 66:2108--2111, 1991.

\bibitem{Daillant1999}
J.~Daillant, S.~Mora, and A.~Sentenac.
\newblock Diffuse scattering.
\newblock In J.~Daillant and A.~Gibaud, editors, {\em X-ray and neutron
  reflectivity: principles and applications, 2nd edition}, pages 133--182.
  Lecture notes in Physics 770, Springer Verlag, Heidelberg, 2009.

\bibitem{Canh70}
P.~Canham.
\newblock {\em J. Theor. Bio.}, 26:61, 1970.

\bibitem{helfrich73}
W.~Helfrich.
\newblock Elastic properties of lipid bilayers: theory and possible
  experiments.
\newblock {\em Zeitschrift f\"ur Naturforschung}, 28:693--703, 1973.

\bibitem{lipoleible1}
R.~Lipowsky and S.~Leibler.
\newblock Unbinding transitions of interacting membranes.
\newblock {\em Physical Review Letters}, 56:2541, 1986.

\bibitem{katsaras}
J.~Katsaras and T.~Gutberlet.
\newblock {\em Lipid Bilayers}.
\newblock Biological Physics Series. Springer, 2000.

\bibitem{israelachvili}
J.N. Israelachvili.
\newblock {\em Intermolecular and Surface Forces}.
\newblock Academic Press, 1992.

\bibitem{Lipowsky95}
R.~Lipowsky.
\newblock In R.~Lipowsky and E.~Sackmann, editors, {\em Handbook of biological
  physics}, page 521. Elsevier, 1995.

\bibitem{Marcelja1976129}
S.~Marcelja and N.~Radic.
\newblock Repulsion of interfaces due to boundary water.
\newblock {\em Chemical Physics Letters}, 42(1):129 -- 130, 1976.

\bibitem{parsegian(lang1991)}
V.~Adrian. Parsegian and R.~Peter. Rand.
\newblock On molecular protrusion as the source of hydration forces.
\newblock {\em Langmuir}, 7(6):1299--1301, 1991.

\bibitem{israelJPhysChem92}
J.~N. Israelachvili and H.~Wennerstrom.
\newblock Entropic forces between amphiphilic surfaces in liquids.
\newblock {\em Journal of Physical Chemistry}, 96:520--531, 1992.

\bibitem{besseling(langmuir1997)}
N.~A.~M. Besseling.
\newblock Theory of hydration forces between surfaces.
\newblock {\em Langmuir}, 13(7):2113--2122, 1997.

\bibitem{helfrich78}
W.~Helfrich.
\newblock Steric interaction of fluid membranes in multilayer systems.
\newblock {\em Zeitschrift f\"ur Naturforschung}, 33:305--315, 1978.

\bibitem{seifert(prl1995)}
U.~Seifert.
\newblock Self-consistent theory of bound vesicles.
\newblock {\em Physical Review Letters}, 74:5060--5063, 1995.

\bibitem{Podgornik1992}
R.~Podgornik and V.A. Parsegian.
\newblock Thermal-mechanical fluctuations of fluid membranes in confined
  geometries: the case of soft confinement.
\newblock {\em Langmuir}, 8:557--562, 1992.

\bibitem{Sinha88}
S.~K. Sinha, E.~B. Sirota, and S.~Garoff.
\newblock X-ray and neutron scattering from rough surfaces.
\newblock {\em Phys. Rev. B}, 38(4):2297--2311, 1988.

\bibitem{daillantrop2000}
J.~Daillant and M.~Alba.
\newblock High-resolution x-ray scattering measurements: I surfaces.
\newblock {\em Reports on Progress in Physics}, 63:1725--1777, 2000.

\bibitem{thesemora}
S.~Mora.
\newblock {\em Structure d'interface de fluides complexes}.
\newblock PhD thesis, Universit\'e Paris 11, 2003.

\bibitem{Nagle89}
J.F. Nagle and Wiener F.C.
\newblock {Relations for lipid bilayers - connection of electron density
  profiles to other structural quantities}.
\newblock {\em {Biophysical Journal}}, {55}({2}):{309--313}, {FEB} {1989}.

\bibitem{Wiener89}
M.C. Wiener, R.M. Suter, and J.F. Nagle.
\newblock Structure of the fully hydrated gel phase of
  dipalmitoylphosphatidylcholine.
\newblock {\em Biophysical Journal}, 55:315--25, 1989.

\bibitem{evans(jphyschem1987)}
E.~Evans and D.~Needham.
\newblock Physical properties of surfactant bilayer membranes: thermal
  transitions, elasticity, rigidity, cohesion, and colloidal interactions.
\newblock {\em Journal of Physical Chemistry}, 91:4219--4228, 1987.

\bibitem{Lipowsky1993}
R.~Lipowsky and S.~Grotehans.
\newblock Hydration vs. protrusion forces between lipid bilayers.
\newblock {\em Europhys. Lett.}, 23:599--604, 1993.

\bibitem{lindahl(biophys2000)}
E.~Lindahl and 0.~Edholm.
\newblock Mesoscopic undulations and thickness fluctuations in lipid bilayers
  from molecular dynamics simulations.
\newblock {\em Biophysical Journal}, 79:426--433, 2000.

\bibitem{leslie09}
I.~Solomonov, K.~Kjaer, J.S. Micha, F.~Rieutord, G.~Fragneto, J.~Daillant, and
  L.~Leiserowitz.
\newblock Cholesterol:phospholipid bilayer membranes at the solid-water
  interface.
\newblock {\em Eur. Phys. J. E, to be published}, 2009.

\bibitem{numrec}
W.H. Press, S.A. Teukolsky, W.T. Vetterling, and B.P. Flannery.
\newblock {\em Numerical recipes in C}, chapter Evaluation of functions.
\newblock Cambridge University Press, 1999.

\end{thebibliography}

\end{document}